\documentclass[11pt]{article}
\pdfoutput=1 
\usepackage{jheppub} 
\usepackage[T1]{fontenc}

\def\bk{\bar{\kappa}}
\def\bc{\bar{c}}

\newcommand{\bea}{\begin{eqnarray}}
\newcommand{\eea}{\end{eqnarray}}
\newcommand{\be}{\begin{equation}}
\newcommand{\ee}{\end{equation}}
\newcommand{\beq}{\begin{equation}}
\newcommand{\eeq}{\end{equation}}

\def\({\left(}
\def\){\right)}
\def\[{\left[}
\def\]{\right]}

\def\half{\frac{1}{2}}

\def\tq{\tilde{q}}
\def\tm{\tilde{m}}

\def\CA{{\cal A}}

\def\CC{{\cal C}}

\def\CE{{\cal E}}

\def\CM{{\cal M}}

\def\CW{{\cal W}}

\definecolor{palatinatepurple}{rgb}{0.41, 0.16, 0.38}

\definecolor{uglybrown}{rgb}{0.8,  0.7,  0.5}

\def\bk{\bar{k}}
\def\ee{\epsilon}
\def\eu{\epsilon_{u}}
\def\ev{\epsilon_{v}}

\title{Subsystem Complexity and Holography}

\author[a.b]{Cesar A. Ag\'on,}
\author[c,d]{Matthew Headrick,}
\author[e]{and Brian Swingle}

\affiliation[a]{C.N. Yang Institute for Theoretical Physics, State University of New York, Stony Brook NY 11794, USA}
\affiliation[b]{Kavli Institute for Theoretical Physics, University of California, Santa Barbara CA 93106, USA}
\affiliation[c]{Martin Fisher School of Physics, Brandeis University, Waltham MA 02453, USA}
\affiliation[d]{Center for Theoretical Physics, Massachusetts Institute of Technology, Cambridge MA 02140, USA}
\affiliation[e]{Condensed Matter Theory Center, Maryland Center for Fundamental Physics,
Joint Center for Quantum Information and Computer Science,
and Department of Physics, University of Maryland, College Park MD 20742, USA}

\abstract{
As a probe of circuit complexity in holographic field theories, we study subsystem analogues based on the entanglement wedge of the bulk quantities appearing in the ``complexity = volume'' and ``complexity = action'' conjectures. We calculate these quantities for one exterior region of an eternal static neutral or charged black hole in general dimensions, dual to a thermal state on one boundary with or without chemical potential respectively, as well as for a shock wave geometry. We then define several analogues of circuit complexity for mixed states, and use tensor networks to gain intuition about them. In the action approach, we find two possible cases depending on an ambiguity in the definition of the action associated with a counterterm. In one case, there is a promising qualitative match between the holographic action and what we call the \emph{purification complexity}, the minimum number of gates required to prepare an arbitrary purification of the given mixed state. In the other case, the match is to what we call the \emph{basis complexity}, the minimum number of gates required to prepare the given mixed state starting from a minimal complexity state with the same eigenvalue spectrum. One way to fix this ambiguity is to choose an action definition such that UV divergent part is positive, in which case the best match to the action result is the basis complexity. In contrast, the holographic volume does not appear to match any of our definitions of mixed-state complexity.

}

\preprint{BRX-TH-6331, YITP-18-02,  MIT-CTP/4989}

\begin{document}
\maketitle
\flushbottom

\section{Introduction}

There has been much recent progress in understanding how spacetime emerges from field theory degrees of freedom within the AdS/CFT correspondence. Considerations involving entanglement \cite{Maldacena_EBH_ADS,Ryu_RT,Raamsdonk_GravityEntanglement,Swingle:2014uza,Lashkari:2013koa,Faulkner:2013ica}, quantum error correction~\cite{Harlow_AdS_QEC}, and other ideas from quantum information science have provided new clues concerning the emergence of the classical bulk geometry as well as the reconstruction of approximately local quantum fields in the bulk \cite{Pastawski:2015qua,Harlow_RT_QEC,Molina-Vilaplana_Hartman-Maldacena,Hayden_RandomTensors}.

Tensor networks provide one set of toy models that instantiate many of the features of AdS/CFT~\cite{Swingle_AdS_MERA} and that can also describe in detail the physics of more conventional systems \cite{Evenbly_BranchingMERA_Entanglement,Swingle_AreaLawEntanglementThermo}. Motivated by these tensor network models and by considerations involving the dynamics of black hole interiors, it was proposed that the quantum computational complexity of the boundary field theory state would also be encoded geometrically in the dual gravitational spacetime~\cite{Susskind_Complexity,Stanford_ComplexityShocks,Brown_CA&BH,Chapman_Complexity_Formation}.

To be more specific, in the context of the eternal AdS-Schwarzchild black hole it was observed that the wormhole which connects the two sides grows linearly with time, say as measured by the length of a geodesic stretching through the wormhole \cite{Hartman_BH_interior,Susskind_Complexity}. One can then ask what the CFT dual of this linear growth is. The conjecture is that the growth of the wormhole is dual to the growth of complexity of the dual CFT state. Roughly speaking, the complexity of the CFT state is the minimum number of simple unitaries or ``gates'' needed to prepare the CFT state from a fixed reference state.

The physical picture is that the complexity of a state can increase due to Hamiltonian time evolution, and it is this increase of complexity that is dual to the late-time growth of the interior. Tensor network models again provide a concrete instantiation of complexity on the CFT side, with the complexity being defined as the number of tensors in the minimal network that describes the state. On the field theory side, one of the key open questions is how to define complexity more precisely. On the gravity side, one of the key issues is how to differentiate between different bulk proposals, including ``complexity equals volume'' (CV) \cite{Susskind_Complexity,Stanford_ComplexityShocks}, ``complexity equals action'' (CA) \cite{CA,Brown_CA}, and others \cite{Abt:2017pmf}. There are by now a large number of papers developing and extending these ideas~\cite{Carmi_CommentsHC, Couch_NoetherCV, Huang_HoloC&2identities, Cai_AdSBH_growthrate, Cai_ActionChargedBH,Yang_SEC&Cbound,Fu_HCnonlocal,Reynolds_CindS,Kuang_EMD, HIS_complexity&gauge, Mansoori_CG&BHT,Alishahiha_Complexity&FR, Momeni_fidelity,Ghodrati-CG,An&Peng-HCGDilaton,AFMM-CGLifshitz-HV}.

In this paper, we consider the problem of defining and evaluating complexity for subsystems of the CFT. To define subsystem complexity on the CFT side, we face the problem of defining complexity for mixed states, since subsystems will generically not be in pure states. On the gravity side, we must find suitable geometric measures which combine the global complexity measures with the physics of subsystem duality in AdS/CFT. Holographic subsystem complexity has recently been studied in a variety of works, with several proposals analogous to CV and CA being advanced and studied \cite{2015PhRvD..92l6009A,Carmi:2016wjl,Carmi:2017jqz,Caceres:2018luq,2017PhLB..766...94M,2017arXiv171103125Z}.

This study was motivated by a desire to better understand the relationships between CA and CV duality and to subject the basic idea of a complexity/geometry duality to a new set of tests. We also wanted to gain insight into the way different subregions of the bulk might be represented in tensor network models.

Our main contributions are as follows. First, we study the analogs of CA and CV duality for subsystems for simple subregions of eternal black holes. This study include calculations of actions and volumes for a variety of black holes as a function of dimension, temperature, and charge. Second, we define a variety of measures of mixed state complexity and compare our definitions to the CA and CV calculations. This analysis complements and extends various discussions given in the holographic literature. Amongst definitions of mixed state complexity that we consider, we find that CA duality reasonably accords with one of two different definitions depending on an apparently arbitrary choice in the definition of the action,
but that CV duality is harder to consistently reconcile with our notions of subsystem complexity.

To more precisely define the holographic quantities we consider, recall that there are reasons to believe that the reduced density matrix for a spatial region $A$ in a holographic theory is encoded in the corresponding entanglement wedge $\CE_A$ \cite{CzechHoloDM,HeadrickPropHoloEE,Headrick:2014cta,Harlow_AdS_QEC, Dong:2016eik}. Combining this observation with the CV and CA proposals for holographic complexity leads one to consider two bulk quantities that can be defined for a given region $A$: the volume $\CC^{\rm V}(A)$ of a maximal Cauchy slice for $\CE_A$ anchored on $A$; and the action $\CC^{\rm A}(A)$ of the Wheeler-de Witt (WdW) patch $\CW_A$ of $\CE_A$ associated to $A$.\footnote{These prescriptions for holographic subsystem complexity were first suggested (to our knowledge) in \cite{2015PhRvD..92l6009A} and \cite{Carmi:2016wjl} for $\CC^{\rm V}(A)$ and $\CC^{\rm A}(A)$ respectively} $\CW_A$ is defined as the set of points in $\CE_A$ that are spacelike- or null-related to $A$ (i.e.\ not in $I^+(A)\cup I^-(A)$), or equivalently as the intersection of $\CE_A$ with the WdW patch $\CW$ of any complete boundary slice containing $A$.\footnote{The equivalence between these two definitions of subsystem WdW patch can be shown as follows. The first definition is $\CW_A^1:=\CE_A\setminus(I^+(A)\cup I^-(A))$. Let $\sigma$ be a boundary Cauchy slice containing $A$. Then $\CW$ is the complement of $I^+(\sigma)\cup I^-(\sigma)$, so the second definition is $\CW_A^2 := \CE_A\setminus(I^+(\sigma)\cup I^-(\sigma))$. Clearly $I^\pm(A)\subseteq I^\pm(\sigma)$, so $\CW_A^1\supseteq\CW_A^2$. For the other direction, let $x$ be a point in $\CW_A^1$. Since $x\in\CE_A$, it is not timelike-related to $\CE_{\sigma\setminus A}$ \cite{Headrick:2014cta}, and in particular is not timelike-related to $\sigma\setminus A$. Since $x$ is also not timelike-related to $A$, it is not timelike-related to $\sigma$, and is therefore in $\CW_A^2$.}

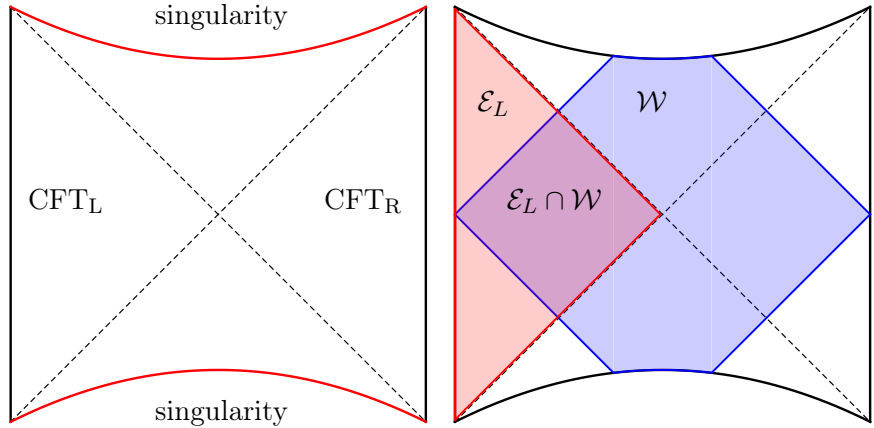
\begin{figure}
$$
\begin{array}{cc}
 \includegraphics[width=2.27in
 ]{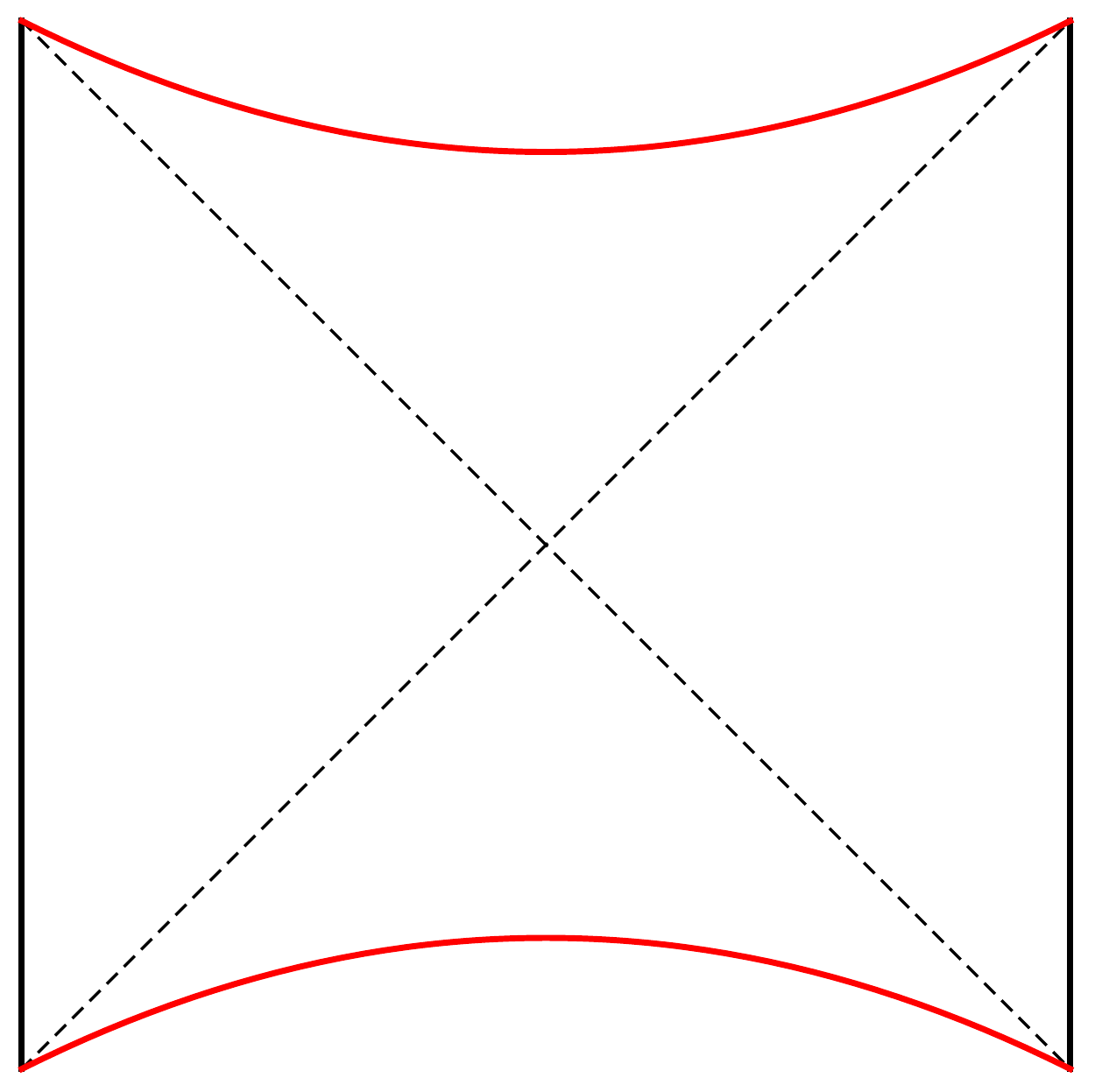} &
  \includegraphics[width=2.27in
  ]{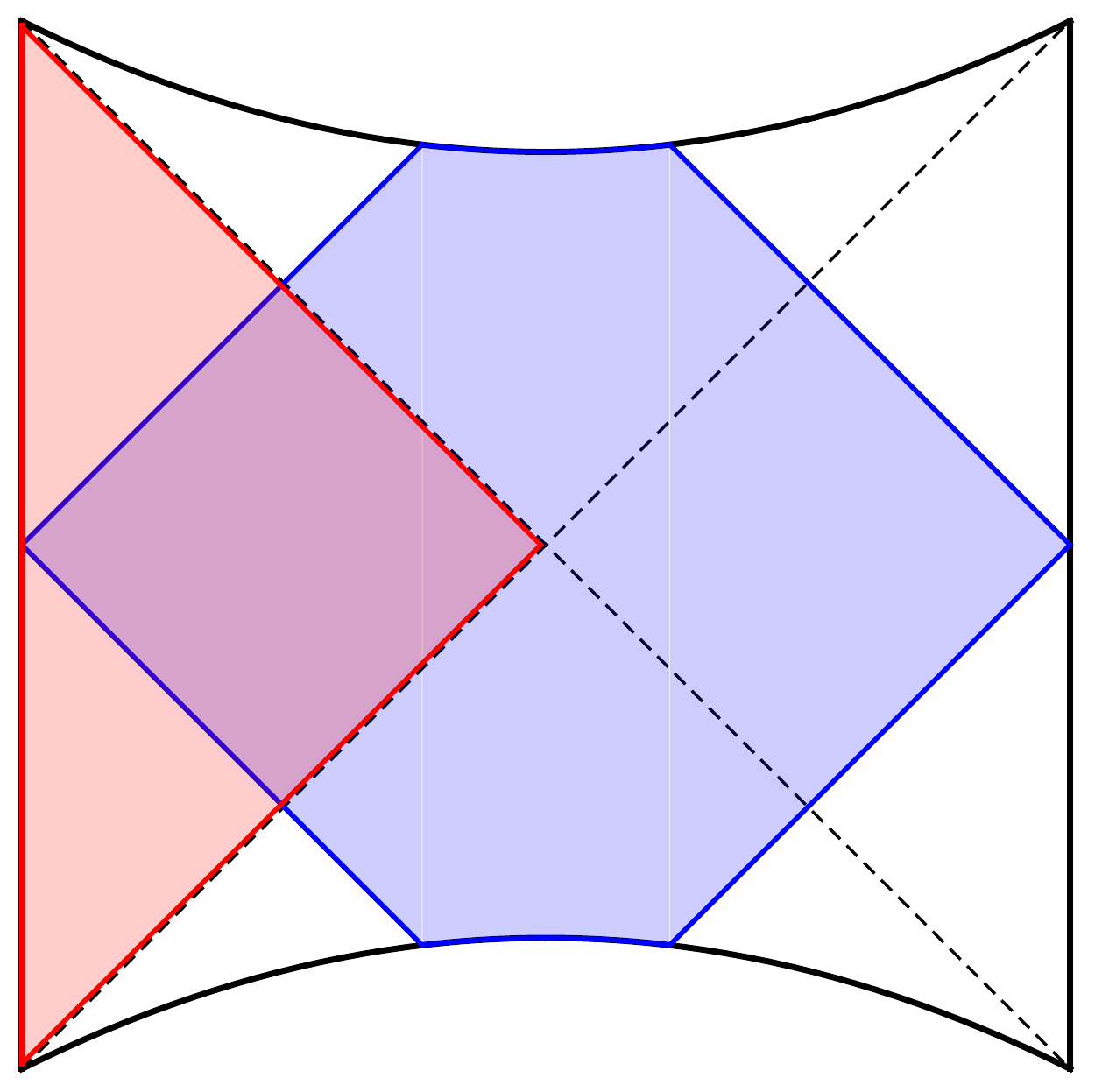}
\end{array}
$$
\begin{picture}(0,0)
\put(113,166){singularity}
\put(65,96){CFT$_{\textrm{L}}$}
\put(177,96){CFT$_{\textrm{R}}$}
\put(113,16){singularity}
 \put(235,133){$\mathcal{E}_L$}
  \put(295,133){$\mathcal{W}$}
   \put(246,96){$\mathcal{E}_L\cap \mathcal{W}$}
 \end{picture}
\caption{Intersection of entanglement wedge $\CE$ and Wheeler-DeWitt patch $\CW$ for half of an eternal black hole.}
\label{fig:eternalbh}
\end{figure}

There are many possible situations in which we could study the above volume and action quantities. For definiteness, in this paper we focus on two-sided (neutral and charged) static AdS black holes, with the subsystem $L$ consisting of a constant-time slice of one boundary. The state of $L$ is thus a Gibbs state (with or without a chemical potential). The entanglement wedge $\CE_L$ of one boundary for a static black hole consists of the corresponding exterior region. Figure \ref{fig:eternalbh} illustrates $\CE_L$, the WdW patch $\CW$ for a complete boundary slice, and their intersection $\CW_L$. In section \ref{sec:holographic}, we calculate $\CC^{\rm A}(L)$ and $\CC^{\rm V}(L)$ for neutral black holes in $D\ge3$ and charged black holes in $D\ge4$. Note that the corresponding Gibbs states are static, reflecting the fact that the bulk isometry generated by the timelike Killing vector of the exterior region relates different constant-time slices of the boundary. Therefore the quantities $\CC^{\rm A}(L)$ and $\CC^{\rm V}(L)$ are time-independent. This is in contrast to the corresponding quantities for the full system, whose state, the thermofield double, is not static.
We also compute $\CC^{\rm A}(L)$ for a thermalizing system dual to a shock wave geometry, finding that at late times it grows linearly at a rate $2M$, just as for the thermofield double.

In section~\ref{sec:measures}, we define various measures of complexity for subsystems as well as for general mixed states (i.e.\ mixed states considered independently of any particular purification). In this discussion, we assume that some notion of pure-state complexity has been previously defined. This notion, however, can be used in different ways to define the complexity of a mixed state $\rho$. For example, we can consider a purification of $\rho$ and evaluate its complexity, or we can decompose $\rho$ into an ensemble of pure states and average their complexities. Other choices in the definition include whether to include ancilla degrees of freedom. By estimating the value of each measure on Gibbs states, and in particular its relation to the entropy, and comparing to the results obtained in section \ref{sec:holographic}, we are able to rule out some of the proposed definitions as being related to either $\CC^{\rm A}$ or $\CC^{\rm V}$.
On the other hand, depending on how the action is defined, we find a promising qualitative match between $\CC^{\rm A}$ and either the \emph{purification complexity} or the \emph{basis complexity}. Roughly speaking, these are the minimum complexity of any purification of the given mixed state and the minimum complexity needed to prepare the basis of the given mixed state, respectively.

The appendices contain certain details of the calculations, including a careful treatment of the corner terms that arise in the action calculations.

\section{Holographic calculations}\label{sec:holographic}

In this section, we consider static two-sided asymptotically AdS black holes, and take the region $L$ to be a constant-time slice of one boundary. Its entanglement wedge $\CE_L$ is the corresponding exterior region in the bulk. We compute the volume $\CC^{\rm V}(L)$ of a maximal slice and the action $\CC^{\rm A}(L)$ of the Wheeler-de Witt (WdW) patch for this exterior region. These quantities are relevant for the subsystem analogues of ``complexity equals volume'' (CV) and ``complexity equals action'' (CA) dualities, respectively. We comment on the differences between the two quantities and the possibility that both dualities hold as they could potentially provide information about different notions of subsystem complexity. We treat the neutral case in subsection \ref{eternalBH} and the charged one in subsection \ref{eternalCBH}. We summarize the results at the top of each subsection before entering into the details of the calculations. First, however, in subsection \ref{sec:relation}, we make a qualitative observation concerning $\CC^{\rm V}(L)$ and $\CC^{\rm A}(L)$ that will play an important role when we compare these quantities to candidate complexity measures in Section \ref{sec:measures}.

Finally, in subsection \ref{sec:shock} we compute $\CC^{\rm A}(L)$ for a shockwave geometry, dual to a system undergoing thermalization after an injection of energy. We find exactly the same late-time behavior as for the two-sided black holes, namely a linear growth at a rate $2M$.

\subsection{Relation between subsystem and full-system measures}\label{sec:relation}

We begin with a general observation about the volume measure $\CC^{\rm V}$. Let $\sigma$ be a boundary Cauchy slice, $A$ a region of $\sigma$, and $A^c:=\sigma\setminus A$ its complement. We assume that the full system on $\sigma$ is in a pure state. Then we
claim that
\begin{equation}\label{CVineq}
\CC^{\rm V}(A)+\CC^{\rm V}(A^c) \le \CC^{\rm V}(\sigma)\,,
\end{equation}
in other words $\CC^{\rm V}$ is superadditive.\footnote{This property was noted independently by the authors of \cite{Phuc-2018}} The reason is that the left-hand side equals the maximum volume of a complete Cauchy slice bounded by $\sigma$ which is constrained to pass through the HRT surface $m(A)$ (which is the same as $m(A^c)$, since the full system is pure), while the right-hand side is the maximum volume of a Cauchy slice that is bounded by $\sigma$ but not constrained to pass through $m(A)$.

In the case we deal with in this paper, where $A=L$ is a constant-time slice of one boundary of a static two-sided black hole and $A^c=R$ is a constant-time slice of the other boundary, we have symmetries that allow us to say more. First, $\CC^{\rm V}(L)$ and $\CC^{\rm V}(R)$ are independent of time, and there is an isometry that exchanges the left and right sides, so
\begin{equation}
\CC^{\rm V}(L) = \CC^{\rm V}(R)\,.
\end{equation}
The full system is not static, so $\CC^{\rm V}(\sigma)$ depends on the times chosen for $R$ and $L$. However, if these are both chosen at $t=0$, then the time-reflection isometry of the black-hole is respected, so by symmetry the maximal-volume slice for $\sigma$ must pass through the bifurcation surface, which is also $m(L)$ and $m(R)$. The inequality \eqref{CVineq} is thus saturated:
\begin{equation}\label{fullpartV}
\CC^{\rm V}(\sigma,t=0) = \CC^{\rm V}(L)+\CC^{\rm V}(R) = 2 \,\CC^{\rm V}(L)\,.
\end{equation}

\begin{figure}
$$
 \includegraphics[width=3in, height=3.2in]{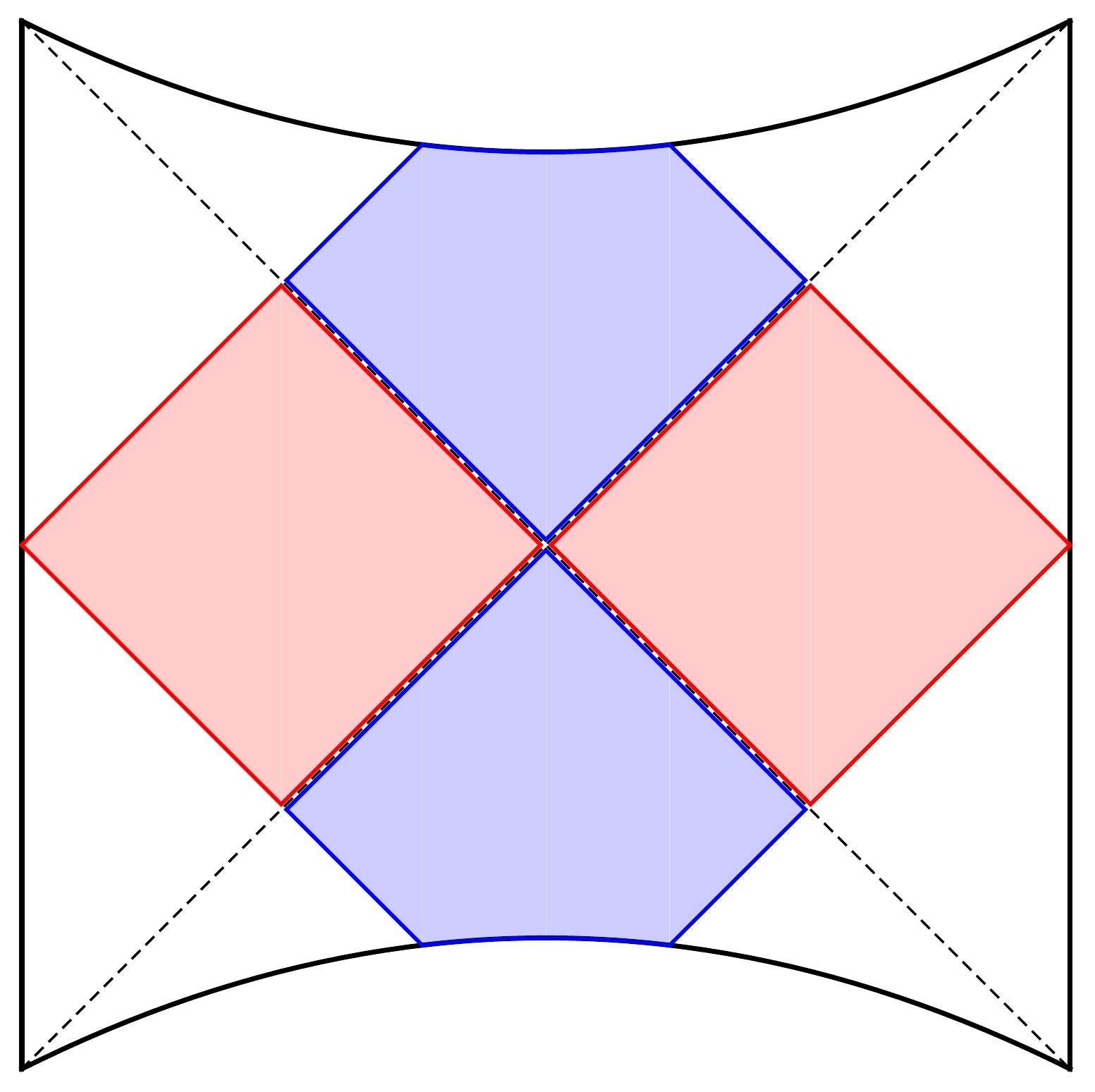}
 $$
\begin{picture}(0,0)
\put(140,130){$\mathcal{E}_L\cap \mathcal{W}$}
\put(250,130){$\mathcal{E}_R\cap \mathcal{W}$}
 \put(210,183){$\mathcal{W}^{+}_{\rm int}$}
  \put(210,80){$\mathcal{W}^{-}_{\rm int}$}
 \end{picture}
\caption{Separation of the WdW patch in terms of its intersection with the entanglement wedges $\CE_L \cap \CW $ and $\CE_R \cap \CW $, and with the regions behind the past and future horizons $\CW_{\rm int}^{\pm}$ for an eternal black hole.}
\label{fig2}
\end{figure}
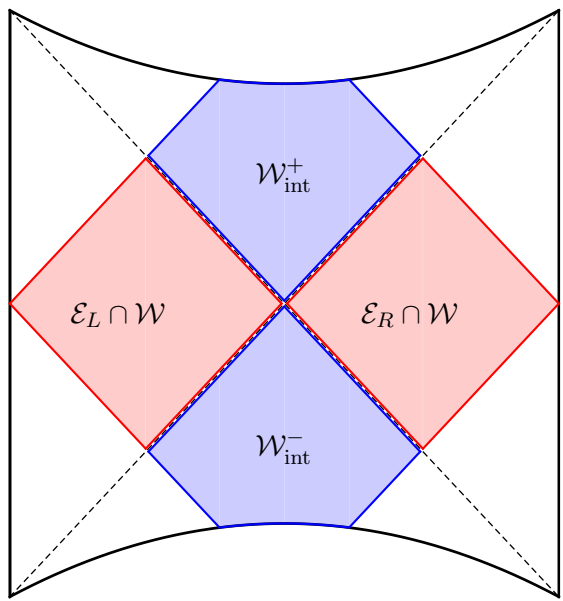

On the other hand, we find that the additivity property of $\CC^{\rm A}$ depends on the definition of the action. As detailed below, there is a counterterm associated with the null surfaces bounding the WDW patch which is necessary to render the action reparameterization invariant, and this counterterm comes with an arbitrary length scale in its definition.\footnote{The effects of the counterterm on subsystem CA duality have also been considered in Ref.~\cite{2018arXiv180906031A}. They define another notion of subsystem complexity on the holographic side obtained from a part of the entanglement wedge which would be interesting to understand further.} As this length scale is varied, the zero time $\CC^{\rm A}$ result goes from being subadditive,
\begin{equation}\label{CAineq}
\CC^{\rm A}(L)+\CC^{\rm A}(R) \geq \CC^{\rm A}(\sigma)\,,
\end{equation}
to being superadditive. Unlike $\CC^{\rm V}$, it will not generically be the case that the action is exactly additive due to the regions behind the horizon. One way to fix this ambiguity in $\CC^{\rm A}$ is to demand that the UV divergent part of the complexity be positive; in this case, the action is superadditive at time zero,
\begin{equation}\label{CAineq}
\CC^{\rm A}(L)+\CC^{\rm A}(R) \leq \CC^{\rm A}(\sigma)\,.
\end{equation}

In more detail, while in the $\CC^{\rm V}$ computation one deals with positive-definite quantities, in the action calculation we have different contributions whose sign depends on the gravitational Lagrangian as well as the boundary and corner terms. For the cases we consider, the Lagrangian is negative, and even including boundary and corner contributions the different regions into which one can decompose the action calculation all give negative results. More precisely, the calculation of the pure state action $\CC^{\rm A}(\sigma)$ can be decomposed as
 \bea\label{fullpart}
 \CC^{\rm A}(\sigma)=\CC^{\rm A}(L)+\CC^{\rm A}(R)+\CA_{\rm int}^+ +\CA_{\rm int}^-\,,
 \eea
where $\CA^{\pm}_{\rm int}$ corresponds to the total action associated to the spacetime region $\CW^{\pm}_{\rm int}$ defined as the intersection of the future/past interior of the black hole with the WdW patch as shown in Figure \ref{fig2}.For example, for the neutral black hole at $t=0$, one has the relation
\begin{equation}
\CC^{\rm A}(\sigma,t=0) =\CC^{\rm A}(L)+\CC^{\rm A}(R)
 +\frac{2S}\pi g_D\,,
\end{equation}
where $S$ is the entropy, $D$ is the bulk spacetime dimension, and $g_D$ is the following $D$-dependent constant:
\bea
g_D=\log\(\frac{l_c}{\ell} (D-2)\)+\frac12 \left[\psi_0(1)-\psi_0\(\frac{1}{D-1}\) \right]+\(\frac{D-2}{D-1}\)\pi
\eea
where $\psi_0(z)$ is the digamma function given by $\psi_0(z)=\Gamma'(z)/\Gamma(z)$. Here $l_c/\ell$ is an undetermined parameter appearing in the definition of the action,
as explained in section \ref{actioncal}. For $l_c/\ell>1$, $g_D$ is positive, but one can find also values of $l_c/\ell<1$ such that $g_D$ is negative. Depending on this choice, the subsystem complexity can be either superadditive or subadditive. This result can be found in (\ref{actioninterior}) of appendix \ref{TFD}. The analogous one for charged black holes can be read from (\ref{purecomplexitycharged}).

\subsection{Neutral black hole}
\label{eternalBH}

Consider an eternal AdS black hole in $D$ spacetime dimensions. In the planar limit, its metric can be written as
\bea
ds^2=\frac{\ell^2}{z^2}\(-f(z)dt^2+\frac{dz^2}{f(z)}+\ell^2 dx^2_{D-2}\)
\eea
where
\beq
f(z) = 1 - \left(\frac{z}{z_h}\right)^{D-1}\,,
\eeq
and the coordinates $(t,x^\mu,z)$ cover one exterior region (left or right)
of the full double-sided geometry, as schematically depicted in Figure \ref{fig:eternalbh}. As argued originally by Maldacena \cite{Maldacena:2001kr}, this geometry is dual to the thermofield double state
\bea
| \psi \rangle\equiv \frac1{Z^{1/2}(\beta)}\sum_{n} e^{-\beta E_{n}/2}|E_{n}\rangle_{L}|E_{n}\rangle_{R}
\eea
where the parameter $\beta$ corresponds to the inverse temperature associated to the eternal black hole; in this case $\beta^{-1}=T_{BH}=(D-1)/4\pi z_{h}$. If one considers expectation values of operators on a single CFT, then the effective state with respect to the CFT$_{L/R}$ is described by a thermal density matrix with inverse temperature $\beta$.
Its entropy and mass are given by
\begin{equation}
S = \frac{V_\perp\ell^{2(D-2)}}{4G_Nz_h^{D-2}}, \qquad \textrm{and }\qquad  M=\frac{V_\perp\ell^{2(D-2)}(D-2)}{16\pi  G_Nz_h^{D-1}}
\end{equation}
respectively, where $V_\perp$ is the dimensionless transverse volume parametrized by the coordinates $x^i$. These quantities equal the entropy and energy of the dual quantum field theory respectively.

We start by computing the WdW action $\CC^{\rm A}(L)$ in subsection \ref{actioncal}. This is done by evaluating the action associated to the region defined as the intersection of the entanglement wedge and WdW patch, including surface and corner contributions as specified by Myers et al.\ \cite{actionnull}.  The result of the thermal state complexity turns out to be remarkably simple:
\bea\label{complexT}
\CC^{\rm A}(L)=\frac{ V_{\perp} \ell^{2(D-2)}}{4\pi G_N} \log\(\frac{l_c}{\ell}(D-2) \)\[ \frac{1}{\delta^{D-2}}-\frac{1}{z_h^{D-2}}\]  +\frac{ V_{\perp} \ell^{2(D-2)}}{4\pi G_N z_h^{D-2}}  g_0.
\eea
where $\delta$ is a UV cutoff and the logarithmic term depends on a particular choice of an undetermined parameter in the computation of the corner contributions, as explained in that section.
The $D$-dependent constant $g_0$ is defined in \eqref{g0def}.
The upper index A refers to the fact that this expression is obtained using the CA duality. We can see from (\ref{complexT}) that, apart from the logarithmic term, the thermal state complexity has a simple
relation to the entropy. More precisely, in terms of the boundary quantities (\ref{complexT}) is
\bea\label{bdry-answer}
\CC^{\rm A}(L)= a(D) \(\frac{V}{\delta^{D-2}} \)c_{\rm eff}\, -b(D)S+\frac{g_0}{\pi} S
\eea
where both $a(D)$ and $b(D)$ are positive coefficients given by
\bea\label{aDbD}
a(D)=4 \log\(\frac{l_c}{\ell}(D-2) \) \quad \textrm{and} \quad b(D)= \frac{1}{\pi} \log\(\frac{l_c}{\ell}(D-2) \) \,.
\eea
In (\ref{bdry-answer}), $V=V_{\perp} \ell^{D-2}$ is the dimension full volume of the boundary theory and $c_{\rm eff}=\ell^{{D-2}}/(16 \pi G_{N})$ characterizes the effective number of degrees of freedom of the dual CFT. 
In subsection  \ref{actionvol} we  evaluate the maximum volume $\CC^{\rm V}(L)$,
again obtaining a very simple answer in terms of the black hole horizon, the UV cut off $\delta$, and the number of spacetime dimensions:
\bea\label{volume-complexity1}
{\cal C}^{\rm V}(L)\approx \frac{V_\perp \ell^{2D-3}}{G_N (D-2) \xi } \( \frac{1}{\delta^{D-2}} +\frac{(D-3)}{2}\frac{\sqrt{\pi}\Gamma(\frac{D}{D-1})}{\Gamma(\frac{D+1}{2(D-1)})} \frac{1}{z^{D-2}_h}\)\,,
\eea
where $\xi$ is a length scale required to make the complexity dimensionless. In terms of the boundary quantities this is
\bea
{\cal C}^{\rm V}(L)\approx \tilde{a}(D) \( \frac{\ell}{\xi}\) \(\frac{V}{\delta^{D-2}} \)c_{eff} +\tilde{b}(D) \(\frac{\ell}{\xi}\)S
\eea
where $\tilde{a}(D)$ and $\tilde{b}(D)$ are also positive and given by
\bea\label{aDbD}
\tilde{a}(D)=\frac{16 \pi }{D-2}\quad  \quad \textrm{and} \quad \quad  \tilde{b}(D)= \frac{1}{2}\frac{(D-3)}{(D-2)}\frac{\sqrt{\pi}\Gamma(\frac{D}{D-1})}{\Gamma(\frac{D+1}{2(D-1)})} \,.\nonumber \\
\eea
An interesting feature of this result is the entropy independence for $D=3$.

\subsubsection{$\CC^{\rm A}(L)$\label{actioncal}}

\begin{figure}
  \centering
  \includegraphics[width=\textwidth]{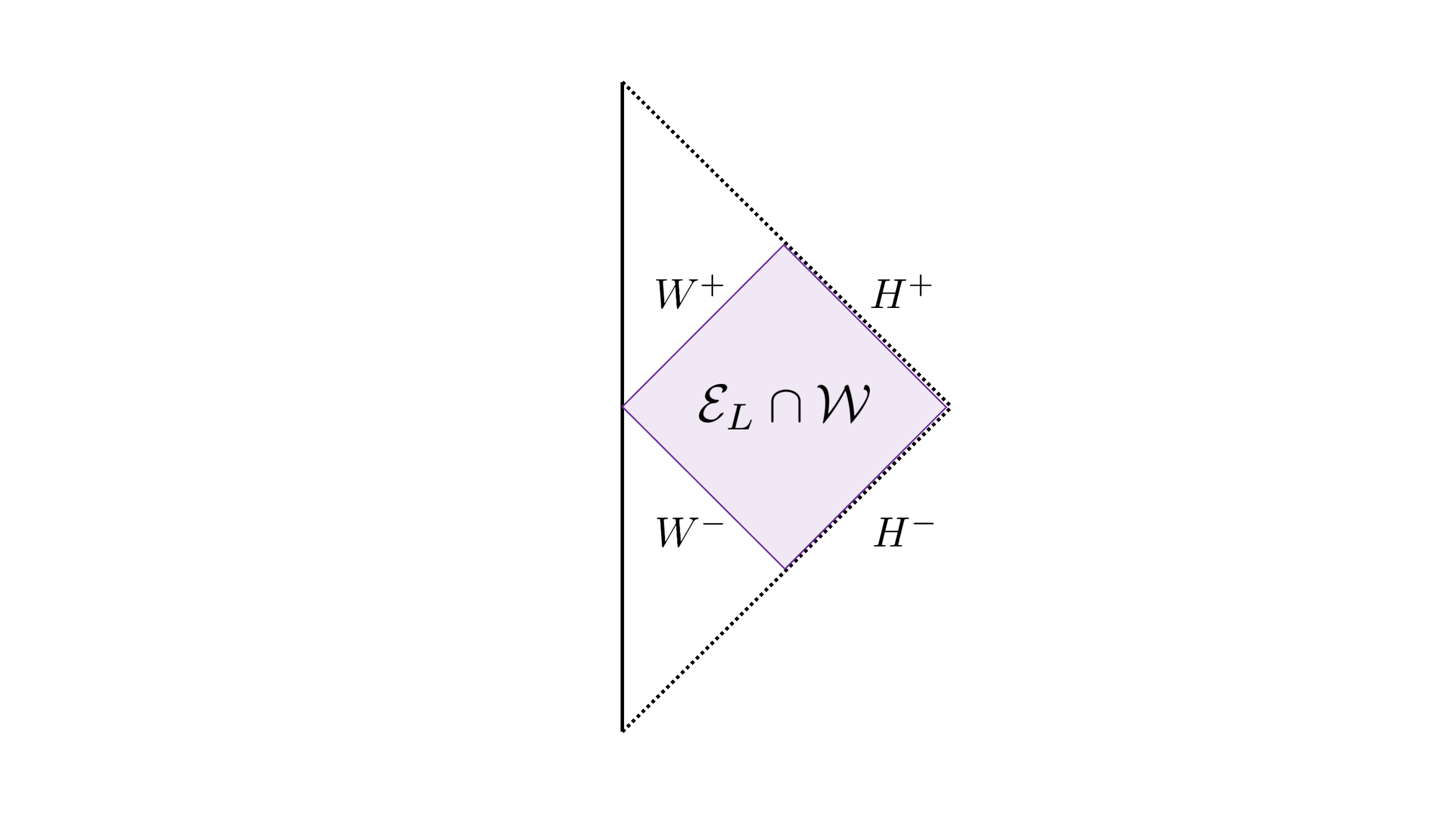}
  \caption{Null segments on the boundary of $\CW_L = \CE_L \cap \CW$. The four corners arise at the intersection of neighboring segments of the boundary. }\label{eternalbh_corners}
\end{figure}

According to the CA duality, the complexity associated to the thermal state describing the left (right) system is given by the action evaluated on the space-time region given by $\CW_{L/R}=\CE_{L/R} \cap \CW$ as ilustrated in Figure  \ref{fig:eternalbh}. In Figure~\ref{eternalbh_corners} we show the intersection of $\CE_L$ and $\CW$ with the null boundaries labelled. $W^\pm$ correspond to boundaries of $\CW$ while $H^\pm$ correspond to the boundaries of $\CE$ and coincide with the black hole horizon.

The action of $\CW_L$ receives three kinds of contributions, one from the bulk, one from the boundaries, and one from the corners \cite{CA,actionnull, Carmi:2016wjl}. We evaluate those following the rules laid out in Ref.~\cite{actionnull}, including the extra counterterms on null boundaries recently discussed in \cite{actionnull,Reynolds:2016rvl} to guarantee the diffeomorphism invariance of the contributions of the null boundary terms.\footnote{We thank the authors of \cite{Alishahiha:2018lfv} for pointing out the relevance of this term in the evaluation of subregion complexity.} The action diverges unless a cutoff is placed near $z=0$. The regulated $\CW$ is defined by starting the null lines that bound $\CW$ from the cutoff surface $z=\delta$.

A convenient set of coordinates that naturally cover the region in question
is obtained by changing $z \to z^*$, where $z^*$ is the tortoise coordinate
\beq
dz^* = \frac{dz}{f(z)}\,,
\eeq
and then defining the light-cone coordinates $u=t-z^*(z)$ and $v=t+z^*(z)$, which can be used to construct the Penrose diagram of Figure \ref{fig:eternalbh}. In these coordinates the metric is
\beq\label{tortoisemetric}
ds^2 = \frac{\ell^2}{z^2} \left[f(z) (-dt^2 + (dz^*)^2) + \ell^2 dx_{D-2}^2 \right].
\eeq
For this family of spacetimes the form of the function $f(z)$ allows an explicit evaluation of the tortoise coordinates:
\bea
z^*(z)&=&z^*(\delta)+ z_h \int^{z/z_h}_{\delta/z_h}\frac{dx}{1-x^{D-1}}\nonumber \\
&=&z^*(\delta)+\frac{z_h}{D-1}\[ B\(x^{D-1};\frac 1{D-1},0\)- B\(\delta_h^{D-1};\frac 1{D-1},0\)\]\label{tortD2}
\eea
where, in the second line, $x\equiv z/z_h$, $\delta_h\equiv \delta/z_h$, and $B(z;a,b)$ is the incomplete beta function given by
\bea \label{in-beta}
B(z;a,b)\equiv \int_0^z u^{a-1} (1-u)^{b-1} du\,.
\eea
We will see in the next section that for the computation we have in mind it is not necesary to have an explicit expression for this function.

First consider the bulk contribution which arises from the bulk action
\beq
\CA_{\text{bulk}} = \frac{1}{16\pi G_N} \int  \sqrt{|g|} (R - 2 \Lambda)
\eeq
where
\begin{equation}
\Lambda=-\frac{(D-1)(D-2)}{2 \ell^2}\,.
\end{equation}
The vacuum Einstein's equations are
\beq
R_{a b} - \frac{R}{2} g_{ab} + \Lambda g_{ab} = 0\,.
\eeq
which leads to $R = \frac{2D}{D-2}\Lambda=-D(D-1)/\ell^2$.
The bulk action is then proportional to the spacetime volume $|{\cal W}_L|$,
\bea
{\cal A}_{ L, {\rm bulk}}=-\frac{(D-1) |{\cal W}_L|}{8\pi G_N \ell^2}\,.
\eea

The spacetime volume is computed as follows. We need the region between the null lines $t(z) = \pm
(z^*(z)-z^*(\delta))$ from $z=\delta$ (a UV regulator) to $z=z_h$, which is
\bea\label{W2}
|{\cal W}_L|&=& 2 V_\perp \ell^{2(D-1)}\int_\delta^{z_h}\frac{dz}{z^D}\( z^{*}(z)-z^*(\delta)\) \nonumber \\
&=&\frac{ 2\,V_\perp \ell^{2(D-1)}}{z_h^{D-2}(D-1)} \int_{\delta_h}^{1} \frac{ dx}{x^{D}} \[ B\(x^{D-1};\frac{1}{D-1},0\)-B\(\delta_h^{D-1};\frac{1}{D-1},0\)\] 
.
\eea
After the change of variable $u\to x^{D-1}$ the integral in (\ref{W2}) can be put in the form
\bea\label{Integral1}
|{\cal W}_L|&=&\frac{ 2\,V_\perp \ell^{2(D-1)}}{z_h^{D-2}(D-1)^2} \int_{\delta_h^{D-1}}^{1} \frac{ du}{u^2} \[ B\(u;\frac{1}{D-1},0\)-B\(\delta_h^{D-1};\frac{1}{D-1},0\)\]. 
\eea
 The remaining integral of the incomplete beta function was computed in Apppendix \ref{A} in equation (\ref{exteriorint}). Using that result, we get for the action
\bea
|{\cal W}_L|=\frac{2 V_\perp \ell^{2(D-1)} }{(D-1)(D-2)}\(\frac{1}{\delta^{D-2}}-\frac{1}{z_h^{D-2}} \)\,,
\eea
so the bulk contribution of the complexity is
\bea\label{bulk-cont}
{\cal A}_{ L, {\rm bulk}}=-\frac{ V_\perp \ell^{2(D-2)}}{4\pi G_N (D-2)} \(\frac{1}{\delta^{D-2}}-\frac{1}{z_h^{D-2}} \)
\eea
up to order $\delta/z_h$.

Now consider the contribution coming from the light sheets which bound the WdW patch. As shown in
\cite{actionnull,Reynolds:2016rvl} such contributions have two pieces for each null hypersurface $N=\{W^{\pm}, H^{\pm}\}$, this is
\bea\label{boundary-ct}
\CA_{\text{boundary}}= \frac{1}{8\pi G_N} \textrm{sgn}(N)\int_{N} d\lambda d^{D-2}x \sqrt{\gamma} \kappa+ \frac{1}{8 \pi G_N} \textrm{sgn}(N)\int_{N} d\lambda d^{D-2}x \sqrt{\gamma} \Theta \log\(l_c |\Theta | \) \nonumber \\
\eea
where the sgn$(N)=1(-1)$ if $N$ lies to the future (past) of the space time region $|\mathcal{W}_L|$, $\kappa$ is the function that appear in the parallel transport of the null generators $k^a$, $ k^a \nabla_a k_b=\kappa k_b$, $\gamma$ is the transverse metric on the null sheet and $\Theta$ is its expansion
\bea
\Theta:=\frac{1}{\sqrt{\gamma}}\partial_\lambda \sqrt{\gamma}\,.
\eea
Notice the undetermined constant $l_c$ in the second term of (\ref{boundary-ct}). Due to the parametrization invariance of (\ref{boundary-ct}), one can evaluate it for any choice of null generators $k^a$. We chose to use affinely parametrized ones for simplicity, which means $\kappa=0$. In this case the first term in (\ref{boundary-ct}) gives zero contribution, and we are left with the so-called counterterm. 
On the horizons $H^\pm$, the expansion vanishes, so the counterterm is zero. 
Meanwhile, the hypersurfaces $W^{\pm}$ are described in Poincar\'e coordinates by $t(z) = \pm
(z^*(z)-z^*(\delta))$. 
We choose to parametrize the generators by $\lambda=-\ell/z$. The transverse metric is simply $\gamma_{ij}=\ell^2 \delta_{ij}/z^2$ and then $\Theta=-(D-2)z/\ell$. Plugging these into (\ref{boundary-ct}) for both $W^+$ and $W^-$ results in
\bea\label{total-boundary}
\CA_{L,\text{boundary}}&=&\frac{2V_{\perp}\ell^{2(D-2)} (D-2)}{8\pi G_N} \int_{\delta}^{z_h} \frac{dz}{z^{D-1}}  \log \(l_c (D-2) z/ \ell \) \nonumber \\
&=& -\frac{V_{\perp}\ell^{2(D-2)}}{4\pi G_N} \[\frac{1}{(D-2)z^{D-2}}+\frac{1}{z^{D-2}}\log \(l_c (D-2) z/\ell \) \]_\delta^{z_h} \nonumber \\
&=&\frac{V_{\perp}\ell^{2(D-2)}}{4\pi G_N(D-2)}\(\frac{1}{\delta^{D-2}}-\frac{1}{z_h^{D-2}} \)+\frac{V_{\perp}\ell^{2(D-2)}}{4\pi G_N}\frac{\log \(l_c (D-2)\delta/\ell \)}{\delta^{D-2}}\nonumber \\
&&-\frac{V_{\perp}\ell^{2(D-2)}}{4\pi G_N}\frac{\log \(l_c (D-2) z_h/\ell \)}{z_h^{D-2}}
\eea
where $V_{\perp}\ell^{D-2}$ comes from the volume integral $\int d^{D-2}x$.
Interestingly, part of the above expression, namely the action contribution from the null boundaries, exactly cancels the full bulk contribution (\ref{bulk-cont}). Notice that in the derivation we did not need to specify the exact form of $f(z)$ and therefore this expression is the same for the charge black hole geometry which we study in the next section.

Now we consider the corner terms, which arise from the action
\beq
\CA_{\text{corner}} = \frac{1}{8 \pi G_N} \int_{\Sigma'} a\, dS
\eeq
where $\Sigma'$ is the corner locus (codimension two) and $a$ is the corner integrand given by
\beq
a = 
\pm \log \left|\frac{k \cdot \bk}{2} \right|
\eeq
where $k,\bk$ are outward directed null normals 1-forms, $k \cdot \bk = g^{ab} k_{a} \bk_{b}$, and 
the sign is determined by the particular corner \cite{actionnull, Carmi:2016wjl}. 
The inner product is easily computed:
\bea
\half k\cdot \bk=-\frac{z^2}{f(z)}
\eea
and therefore for the corner terms we have
\bea
\frac{1}{8\pi G_{N}}\int_{\Sigma'_{i}} a dS=\textrm{sgn}_i \,\frac{ V_\perp}{8\pi G_{N}}\(\frac{\ell^2} {z_i}\)^{D-2}\log\(-\frac{f(z_i)}{z_i^2}\)
\eea
where $z_i$ is the $z$ coordinate at the corner $i$.
Since $f(z)$ is zero at the horizon, one has to be careful when computing the corner contributions there. We do this carefully in Appendix \ref{AA}, where we find that the sum of the four corner terms---one at the boundary, $W^{-} \cap W^{+}$, and three on the horizon, $W^{+} \cap H^{+}$, $H^{+} \cap H^{-}$, and $W^{-} \cap H^{-}$, as illustrated in figure \ref{eternalbh_corners}---is given by
\begin{equation}\label{aiint}
\frac{1}{8\pi G_N}\sum_i\int_{\Sigma_{i}'} a_i dS_i
=\frac{V_{\perp}}{4\pi G_N}\[- \( \frac{\ell^2}{\delta }\)^{D-2} \log\delta +  \( \frac{\ell^2}{z_h}\)^{D-2} (
g_0 + \log z_h )   \],
\end{equation}
where
\begin{equation}\label{g0def}
g_0=\frac12 \left[\psi_0\(\frac{1}{D-1}\)-\psi_0(1) \right]
\end{equation}
and $\psi_0(z)$ is the digamma function given by $\psi_0(z)=\Gamma'(z)/\Gamma(z)$. For  $D>2$, $g_0$ is negative 
(for example, for $D=3$, $g_0=-\log 2$).

The full action is thus
\bea\label{actionbulk}
\CC^{\rm A}_L(T)=\frac{ V_{\perp} \ell^{2(D-2)}}{4\pi G_N} \log\(\frac{l_c}\ell(D-2) \)\[ \frac{1}{\delta^{D-2}}-\frac{1}{z_h^{D-2}}\]  +\frac{ V_{\perp} \ell^{2(D-2)}}{4\pi G_N z_h^{D-2}}  g_0.
\eea
Notice that the arbitrary constants appearing in the boundary and joint contributions form a single dimensionless parameter $l_c/\ell$ in the above formula. This suggests the natural $D$-independent choice $l_c/\ell\geq 1$ since that would be enough to guarantee a positive volume-divergent contribution for the subsystem complexity. However, this also implies that the first correction to the positive divergent term in $\CC^{\rm A}_L(T)$ is negative. As we will see this is an important qualitative feature of this quantity. On the other hand, we might impose the condition that the finite subleading contribution should be positive instead, in which case then one would need to make $\ell/l_c>c'(D-2)$ for some constant $c'$ and then the condition would be dimension-dependent and less natural.

\subsubsection{$\CC^{\rm V}(L)$\label{actionvol}}

In the spirit of the volume equals complexity conjecture, we would like to compute, for the family of geometries we considered in the previous section, the volume of the maximal slice bounded by $\Sigma$ and the HRT surface $m(\Sigma)$. In the case at hand $\Sigma$ is the boundary region $t=0, z=\delta$, and $m(\Sigma)$ is the horizon $z=z_h$ (where the coordinate $t$ is undefined). In this case, due to the staticness of the metric in the exterior region, it is clear that the maximal slice is just the $t=0$ hypersurface.

The extremal volume $V$ is computed by direct integration:
\bea\label{volume}
V&=&V_\perp \ell^{D-2}\int_\delta^{z_h} dz \frac{\ell^{D-1}}{z^{D-1}} \frac{1}{\sqrt{f(z)}} \nonumber \\
&=& \frac{V_\perp \ell^{2D-3}}{z^{D-2}_h}\int_{\delta/z_h}^{1} \frac{dy}{y^{D-1}\sqrt{1-y^{D-1}}} 
\eea
whose leading order value in the $\delta/z_h$ expansion is\footnote{The following result is obtained after expanding the resulting integral
$$\int^{x} \frac{dy}{y^{D-1}\sqrt{1-y^{D-1}}} =  -\frac{1}{D-2}\frac{\sqrt{1-x^{D-1}}}{x^{D-2}} +\frac{(D-3)}{2(D-2)} x \,_2F_1\[\frac 12, \frac 1{D-1},\frac{D}{D-1},x^{D-1}\] $$
for $x=1$ and $x=\delta/z_{h}$, where $\,_2F_1[a,b,c,x]$ is the hypergeometric function.
}
\bea
V&\approx &V_\perp \ell^{2D-3} \[ \frac{(D-3)}{2(D-2)}\frac{\sqrt{\pi}\Gamma(\frac{D}{D-1})}{\Gamma(\frac{D+1}{2(D-1)})} \frac{1}{z^{D-2}_h} + \frac{1}{D-2}\frac{1}{\delta^{D-2}}\]\,,
\eea
so the corresponding volume complexity is
\bea\label{volume-complexity}
{\cal C}^{\rm V}(L)\approx \frac{V_\perp \ell^{2D-3}}{G_N \xi } \[ \frac{(D-3)}{2(D-2)}\frac{\sqrt{\pi}\Gamma(\frac{D}{D-1})}{\Gamma(\frac{D+1}{2(D-1)})} \frac{1}{z^{D-2}_h} + \frac{1}{D-2} \frac{1}{\delta^{D-2}} \]\,,
\eea
where $\xi$ is a length scale required to make the complexity dimensionless.

\subsection{Charged black hole}
\label{eternalCBH}

In this section we repeat the analysis of the eternal black hole of subsection \ref{eternalBH} for the more general family of charged black holes characterized by mass and charge parameters $m,q$ and described by the space time metric
\bea\label{metricz}
ds^2=\frac{l^2}{z^2}\left(
-f(z)dt^2+\frac{dz^2}{f(z)} +l^2dx_{D-2}^2 \right)
\eea
with
\bea\label{fmq}
f(z)=1-m \,z^{D-1} +q^2 \,z^{2(D-2)}
\eea
in Poincar\'e-like coordinates. The spacetimes in this family are duals of the thermofield double state
\bea
| \psi \rangle\equiv \frac1{Z^{1/2}(\beta,\mu)}\sum_{n,m} e^{-\beta( E_{n}+\mu \, Q_m)/2}|E_{n},Q_m\rangle_{L}|E_{n}, -Q_m\rangle_{R}\,.
\eea
We are interested in studying the subsytem complexity associated to the left (right) subsystems obtained after tracing out the degrees of freedom of the right (left) parts of the Hilbert space. In this case the resulting reduced state is given by the density matrix describing a grand canonical ensamble, with temperature $T=\beta^{-1}$ and chemical potential $\mu$.

The metric (\ref{metricz}) is the solution to the classical equations of motion derived from the Einstein-Hilbert action in the presence of an electromagnetic field $F_{\mu \nu}$, this is
\bea\label{EMaction}
{\cal A}_{L,{\rm bulk}}=\frac{1}{16 \pi G_N} \int \sqrt{|g|} \(R-2\Lambda\)-\frac{1}{16 \pi G_N} \int \sqrt{|g|}\, F_{\mu\nu}F^{\mu \nu}.
\eea
The energy momentum tensor sourced by the field strength is given by
\bea
T_{\alpha\beta}=\frac 1{4\pi}\left(  F_\alpha^{\,\,\,\mu} F_{\beta \mu}-\frac 14 g_{\alpha \beta} F_{\mu \nu}F^{\mu \nu}\right)\,.
\eea
where $\Lambda=-\frac{(D-1)(D-2)}{2 \ell^2}$ is the cosmological constant.

The solutions to the classical equations of motion for the metric (under the assumption of flat boundary conditions) and the gauge field are given by \cite{Chamblin:1999tk}
\bea
ds^2=-f(r)dt^2+\frac{dr^2}{f(r)} +r^2dx_{D-2}^2
\eea
where
\bea
f(r)=\frac{r^2}{\ell^2}-\frac{\tilde{m}}{r^{D-3}}+\frac{\tilde{q}^2}{r^{2(D-3)}}
\eea
and
\bea\label{gaugepot}
A_\mu dx^\mu=\sqrt{\frac{(D-2)}{2(D-3)}}\left(\frac \tq{r_h^{D-3}}-\frac \tq{r^{D-3}}\right)dt
\eea
where $\tilde{m}$ and $\tilde{q}$ are related to the ADM mass and charge via
\bea
\tm=\frac{16\pi GM}{(D-2)V_\perp}\,, \quad \quad \tq^2=\frac{8\pi G Q^2}{(D-2)V_\perp}\,.
\eea

The metric (\ref{metricz}) is obtained by changing the radial coordinate $r$ to $z=l^2/r$ where the parameters $m, q^2$ in (\ref{fmq}) are related to $\tilde{m}, \tilde{q}^2$ via
\bea
m=\frac{\tm}{\ell^{2(D-2)}}\,, \quad \quad q^2=\frac{\tq^2}{\ell^{2(2D-5)}}\,.
\eea

The function $f(z)$ has two positive zeros $z_\pm$, where $z_-$ is the smaller one. Since the asymptotic boundary of this metric corresponds to the $z\to 0$ region,
the region outside the horizon corresponds to $z<z_{-}$ which implies that $z_-$ is the horizon radius of this metric. The existence of a horizon at $z=z_{h}=z_-$, $f(z_h)=0$, establishes a useful relation between $z_h$, $q^2$, and $m$:
\bea\label{outerhorizon}
m =z_{h}^{1-D} +q^2 \,z_{h}^{D-3}\,.
\eea

In subsections \ref{actioncalq} and \ref{volumencalq}, we compute the mixed state complexity associated to the finite temperature and finite chemical potential density matrix describing the grand canonical ensemble given by $\rho=Z^{-1}e^{-\beta(H+\mu Q)}$ via the CA and CV duality respectively. The boundary temperature $T$ and chemical potential $\mu$ are given by
\bea\label{T-mu-rels}
T= \frac{(D-1)}{4\pi z_h}\( 1-\(\frac{D-3}{D-2}\)q^2 \,z_{h}^{2(D-2)}\) \quad \textrm{and}\quad \mu=\ell \sqrt{\frac{(D-2)}{2(D-3)}}\,\, q \,z_h^{D-3}\,.
\eea
The expression for $T$ is derived from $T=-f'(z_h)/4\pi$ and  the one for $\mu$ can be deduce from  (\ref{gaugepot}) by taking the $r\to \infty$ limit of $A_{0}$.
For the action calculation performed in (\ref{actioncalq}) we obtained a relatively simple answer
\bea\label{complexTmu}
\CC^{\rm A}(L)=\frac{ V_{\perp} \ell^{2(D-2)}}{4\pi G_N} \log\(\frac{l_c}\ell(D-2) \)\[ \frac{1}{\delta^{D-2}}-\frac{1}{z_h^{D-2}}\]  +\frac{ V_{\perp} \ell^{2(D-2)}}{4\pi G_N z_h^{D-2}}  g(z_h).
\eea
where $g(z_h)$ is defined from the following finite limit
\bea
g(z_h)=\lim_{z\to z_h}\left\{\log\(f(z)\)-f'(z_h)\int_0^z \frac{dz}{f(z)}\right\}\,
\eea
as explained in Appendix \ref{AA}.

In fact this result is very similar to the one for neutral black hole (\ref{complexT}), with the only difference being the term $g(z_h)$ and the explicit dependence of the black hole horizon $z_h$ on the mass and charge parameters $m, q^2$.

In terms of the field theory quantities the subsystem complexity for charged black holes is given by
\bea
\CC^{\rm A}(L)= a(D) \(\frac{V}{\delta^{D-2}} \)c_{\rm eff}\,-b(D)S+g(z_h) S
\eea
where the coefficients $a(D)$ and $b(D)$ are given in (\ref{aDbD}), and $z_h$ is a complicated function of the boundary quantities $T$ and $\mu$ which can in principle be derived from (\ref{T-mu-rels}).

In section (\ref{volumencalq}) we study the analogous quantity, $C^{\rm V}(L)$ as given by the CV duality, obtaining explicit expressions in two different regimes corresponding to near extremal black holes where $m z_h^{D-1}=2(D-2)/(D-3)$ and $q^2 z_h^{2(D-2)}=(D-1)/(D-3)$, and weakly charged black holes where $q^2 z_h^{2(D-2)}\ll 1$.

\subsubsection{$\CC^{\rm A}(L)$\label{actioncalq}}

\begin{figure}
$$
 \includegraphics[width=2in, height=3.6in]{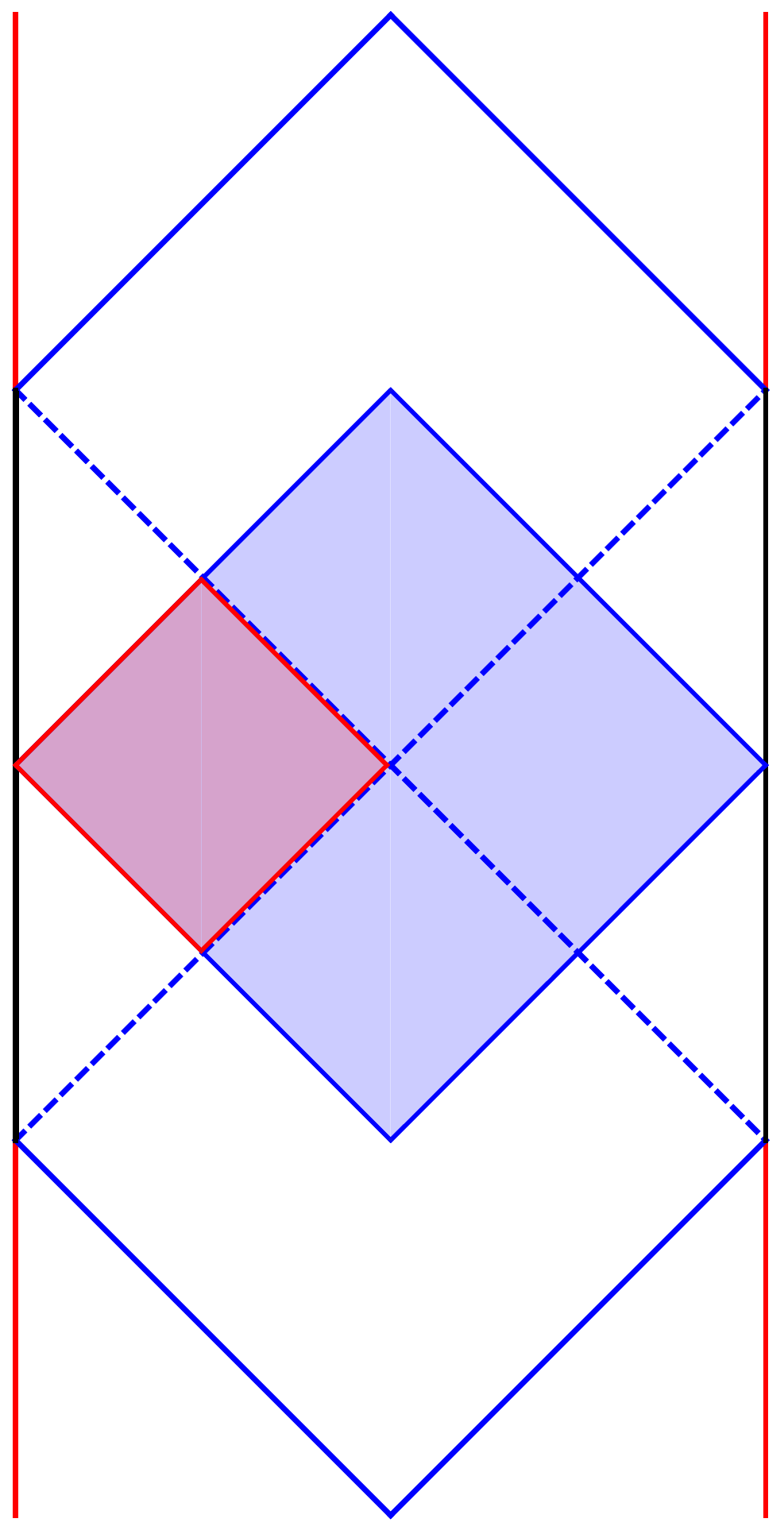}
 $$
 \begin{picture}(0,0)
\put(167,150){\small${\small{\mathcal{E}_L\cap \mathcal{W}}}$}
\put(115,150){CFT$_{\textrm{L}}$}
\put(297,150){CFT$_{\textrm{R}}$}
\put(171,203){${z_h}$} \put(255,203){${z_h}$}
\put(171,95){${z_h}$} \put(255,98){${z_h}$}
\put(120,13){singularity} \put(270,13){singularity}
\put(120,285){singularity} \put(270,285){singularity}
 \end{picture}
\caption{Penrose diagram for an eternal charged black hole. $\CW_L$, the intersection of the entanglement wedge associated to the left boundary $\CE_L$ and the Wheeler-DeWitt patch $\CW$ associated to the full boundary of the black hole, is shown in purple.}
\label{ChargedBH}
\end{figure}
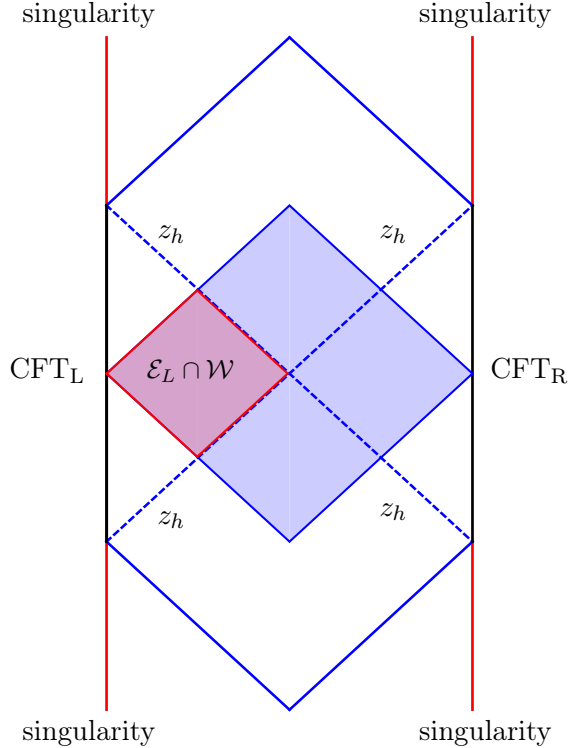

The action evaluation required to compute the subsystem complexity $\CC^{\rm A}$ follows closely the steps laid down in subsection \ref{actioncal}, although the Penrose diagram differs from the uncharged case, as illustrated in Figure \ref{ChargedBH}. The integration region of interest $\CW_L = \CE_L \cap \CW $ is essentially the same, the only difference beging the functions we are integrating. Then, for the bulk evaluation we have
\bea
{\cal A}_{L,{\rm bulk}}=\frac{1}{16 \pi G_N} \int_{\CW_L} \sqrt{|g|} \(R-2\Lambda\)-\frac{1}{16 \pi G_N} \int_{\CW_L} \sqrt{|g|} \,F_{\mu\nu}F^{\mu \nu}\,.
\eea
Taking the trace of the Einstein equation in the presence of the electromagnetic field leads to the relation
\bea\label{RF2}
R=\frac{2D \Lambda}{D-2}+\frac{D-4}{D-2}F_{\mu \nu}F^{\mu \nu}\,.
\eea
The term $F_{\mu \nu}F^{\mu \nu}$ can be evaluated from the gauge field solution (\ref{gaugepot})
\bea
 F_{\mu \nu}F^{\mu \nu}=-2(\partial_i A_0)^2=-{(D-2)(D-3)}\frac{\tq^2}{r^{2(D-2)}}\,.
\eea
This result together with (\ref{RF2}) leads to the following simple expression for the bulk on-shell action
\bea\label{actionbl2}
{\cal A}_{L,{\rm bulk}}=-\frac{(D-1)}{8\pi G_N l^2} \int_{\mathcal{W}_L} \sqrt{|g|} \left(1-\(\frac{D-3}{D-1}\) q^{2}z^{2(D-2)}\right)d^D x\,.
\eea

As in the uncharged case it is convenient to use light cone coordinates $u=t-z^*(z)$, $v=t+z^*(z)$  where $z^*$ is the tortoise coordinate defined as
\beq
dz^* = \frac{dz}{f(z)}\,.
\eeq
These coordinates can be used to construct the Penrose diagram of figure \ref{ChargedBH} and so they naturally cover the region of interest $\CW_L$.

For example, the metric takes the simpler form
\beq\label{tortoisemetric2}
ds^2 = \frac{\ell^2}{z^2} \left[f(z) (-dt^2 + (dz^*)^2) + \ell^2 dx_{D-2}^2 \right],
\eeq
and the light rays which bound the causal region $\CW_L$ are given by $t_{\pm}(z)=\pm(z^*(z)-z^*(\delta))$
where $\delta$ is a UV cut off and
\bea
z^{*}(z)-z^*(\delta)=\int_\delta^{z} \frac{d\xi}{f(\xi)}\,.
\eea

Once the integration region $\mathcal{W}_L$ is delimited, one can explicitly integrate the perpendicular directions since they are independent of it. Doing so leads to a dimensionless volume factor $V_\perp$, and the remaining two-dimensional integral
\bea\label{actionsomething}
{\cal A}_{L,{\rm bulk}}=-\frac{(D-1)V_\perp \ell^{2(D-2)}}{4\pi G_N } \int_{\delta}^{z_{h}} \frac{dz}{z^{D}}
\(1-\(\frac{D-3}{D-1}\) q^{2}z^{2(D-2)}\)\int_\delta^{z} \frac{d\xi}{f(\xi)}\,.
\nonumber \\
\eea

Notice that the $\xi$ integral in (\ref{actionsomething}) is highly non-trivial while the $z$ integral is much simpler. To use this fact in our advantage, consider the integration region in the plane $(z,\xi)$ and invert the order in which we perform the integration. The resulting expression is then
\bea\label{actionsomething2}
{\cal A}_{L,{\rm bulk}}=-\frac{(D-1)V_\perp \ell^{2(D-2)}}{4\pi G_N }\int_\delta^{z_{h}} \frac{d\xi}{f(\xi)} \int_{\xi}^{z_{h}} \frac{dz}{z^{D}}
\(1-\(\frac{D-3}{D-1}\) q^{2}z^{2(D-2)}\)\,.
\nonumber \\
\eea
Let's consider the $z$ integral separately, which evaluates to
\bea
 \int_{\xi}^{z_{h}} \frac{dz}{z^{D}}
\(1-\(\frac{D-3}{D-1}\) q^{2}z^{2(D-2)}\)&=&\frac{\xi^{1-D}}{D-1} \( 1-(z_{h}^{1-D}+q^{2}z_{h}^{D-3}) \xi^{D-1} +q^2 \,\xi^{2(D-2)}\)\nonumber \\
&=& \frac{\xi^{1-D}}{D-1} f(\xi)\,,
\eea
where in the second line we have used the relation between $m, z_{h}$ and $q^{2}$ (\ref{outerhorizon}).
The exact cancellation of the function $f(\xi)$ for arbitrary values of $m$ and $q^{2}$ is remarkable, and leads to the following simple answer for the bulk action
\bea
{\cal A}_{L,{\rm bulk}}=-\frac{V_\perp \ell^{2(D-2)}}{4\pi G_N(D-2)}
\(\frac{1}{\delta^{D-2}} -\frac{1}{z_{h}^{D-2}} \)\,.
\eea
As explained in the previous section, the gravitational boundary contribution\footnote{We assume that the boundary conditions of the electromagnetic fields $F_{\mu \nu}$ are such that they do give rise to boundary terms on the null surfaces} in this case is exactly the same that in the uncharged case and therefore is given by (\ref{total-boundary}). This is
\bea\label{total-boundary-2}
\CA_{L,\text{boundary}}&=&\frac{V_{\perp}\ell^{2(D-2)}}{4\pi G_N(D-2)}\(\frac{1}{\delta^{D-2}}-\frac{1}{z_h^{D-2}} \)+\frac{V_{\perp}\ell^{2(D-2)}}{4\pi G_N}\frac{\log \(l_c (D-2)\delta/\ell \)}{\delta^{D-2}}\nonumber \\
&&-\frac{V_{\perp}\ell^{2(D-2)}}{4\pi G_N}\frac{\log \(l_c (D-2) z_h/\ell \)}{z_h^{D-2}}
\eea
Similarly to what happen to the boundary calculation, the calculation of the corner terms goes exactly as in the section \ref{actioncal}, with the extra details from the Appendix \ref{AA}. This leads to
\begin{equation}
{\cal A}_{L, \text{corner}}=\frac{V_{\perp}}{4\pi G_N}\[ -\( \frac{\ell^2}{\delta }\)^{D-2} \log \delta +  \( \frac{\ell^2}{z_h}\)^{D-2} \left(
g(z) + \log z_h \right)   \] ,
\end{equation}
where $g(z_h)$ is defined from the following limit
\bea
g(z_h)=\lim_{z\to z_h}\left\{\log\(f(z)\)-f'(z_h)\int_0^z \frac{dz}{f(z)}\right\}\,
\eea
which as explained in Appendix \ref{AA} is finite for generic values of $q, m^2$ but has an IR divergence in the extremal case \ref{A3}.

The full action is thus
\bea\label{actionbulk2}
\CC^{\rm A}_L(T)=\frac{ V_{\perp} \ell^{2(D-2)}}{4\pi G_N} \log\(\frac{l_c}\ell(D-2) \)\[ \frac{1}{\delta^{D-2}}-\frac{1}{z_h^{D-2}}\]  +\frac{ V_{\perp} \ell^{2(D-2)}}{4\pi G_N z_h^{D-2}}  g(z_h)\,.
\eea

\subsubsection{$\CC^{\rm V}(L)$\label{volumencalq} }

To evaluate $\CC^{\rm V}(L)$ we need the compute the extremal volume surface which asymptotes to the boundary $t=0$ surface and the bulk minimal surface. The extremal surface is given by the $t=0$ hypersurface in Poincare coordinates, and the extremal volume is therefore given by
\bea\label{volume2}
V&=&V_\perp \ell^{(D-2)}\int_\delta^{z_h} dz \frac{\ell^{D-1}}{z^{D-1}} \frac{1}{\sqrt{f(z)}} \nonumber \\
&=& V_\perp \ell^{2D-3} \int_{\delta}^{z_h} \frac{dz}{z^{D-1}\sqrt{1-m z^{D-1}+q^2 z^{2(D-2)} }}\,. \nonumber \\
\eea
We would like to evaluate this integral for arbitrary values of $m$ and $q^2$ but were unfortunately unable to do so. Nevertheless one can explore the finiteness of the volume answer. Apart from the obvious UV divergence, the presence of poles in $f(z)$ indicate potential divergences in the volume integral. This is easy to do since we know the only real poles of $f(z)$ are at $z=z_{\pm}$ and in the integration region we only encounter the $z=z_-=z_h$ pole, except when $z_\pm$ collide with each other. That is the case of the extremal black holes for which as we will see there is an IR divergence in the volume integral.

Let's study the contribution of the integral in the region close to $z\approx z_h$. First we want
\bea
f(z)\approx f(z_h)+f'(z_h)(z-z_h)+f''(z_h)(z-z_h)^2+\cdots\,.
\eea
On the horizon we have $f(z_h)=0$, therefore if $f'(z_h)\neq 0$ we will have a square root divergence close to $z=z_h$ but a square root divergences integrates to a finite value, this is
\bea
V \supset \int^{z_h}_{z_h-\Delta z} \frac{dz\, }{z^{D-1}\sqrt{-f'(z_h)(z_h-z)}}& \approx & \frac{z_h^{1-D}}{\sqrt{-f'(z_h)}}\int^{z_h}_{z_h-\Delta z} dz (z_h-z)^{-1/2} \nonumber \\
&\approx & -\lim_{z \to z_h}\frac{z_h^{1-D}}{\sqrt{-f'(z_h)}}(z_h-z)^{1/2} +\textrm{finite}\,.\nonumber \\
\eea
That means that for generic values of $m$, $q^2$ the integral is finite.

On the other hand for specific values of $m$, $q^2$ for which $-f'(z_h)\ll 1$ but finite, then, we would have a large contribution from the above integral which scales as
\bea
V\sim \frac{1}{z_h^{D-1}\sqrt{z_h T}}\,.
\eea
This means that $\CC^{\rm V}(L)$, diverges as $1/\sqrt{T}$ as the black hole approaches extremality, since $T=-f'(z_h)/4\pi$.

\paragraph{Extremal black holes:}
However, for $m$, $q^2$ such that $f'(z_h)=0$ we need to go to the next order in the expansion and then
\bea
V \supset \int^{z_h}_{z_h-\Delta z} \frac{dz\, }{z^{D-1}\sqrt{f''(z_h)(z-z_h)^2}}& \approx & -\frac{z_h^{1-D}}{\sqrt{f''(z_h)}}\int^{z_h}_{z_h-\Delta z}\frac{dz}{|z-z_h|} \,,
\eea
which is logarithmically divergent. This is in fact the case for extremal black holes. One can go one step forward and compute the coefficient of the logarithmic IR divergence in $V$ by using the fact that $f(z_h)=f'(z_h)=0$ to show that $f''(z_h) = 2(D-1)(D_2)/z_h^2$:
\bea
V \supset
-\frac{z_h^{2-D}}{\sqrt{2(D-1)(D-2)}}\log\(\frac{\epsilon}{z_h}\) +\textrm{finite}\,,
\eea
where the integral is taken up to $z_h-\epsilon$.
For extremal black holes, we were unable to obtain a closed expression for the volume, although its dependence on $z_h$ can be extracted since both $m$ and $q^2$ parameters in dimensionless units depend only on the number of dimensions $D$:
\bea
m z_h^{D-1}=\frac{2(D-2)}{D-3}, \qquad q^2 z_h^{2(D-2)}=\frac{D-1}{D-3}\,.
\eea
Therefore, the volume of the maximum volume surface in units of $G_N \xi$ (where, again, $\xi$ is a length scale to make $\CC^V$ dimensionless) is given by
\bea
\CC^{\rm V}(L)&=&\frac{V_\perp \ell^{2D-3}}{G_N \xi z_h^{D-2}} \int_{\delta/z_h}^{1-\epsilon_h} \frac{dx}{x^{D-1}\sqrt{1-2\frac{D-2}{D-3}x^{D-1}+\frac{D-1}{D-3}x^{2(D-2)}}} \,.
\eea
The leading divergent term coming from the lower integration point $\delta/z_h$ can be extracted by integrating the region close to $x=0$. This result plus the integration in the region $x=1$ leads to the following structure of the answer:
\bea
\CC^{\rm V}(L)&=&\frac{V_\perp \ell^{2D-3}}{(D-2)G_N \xi \delta^{D-2}} - \frac{V_\perp \ell^{2D-3}}{\sqrt{2(D-1)(D-2)}G_N \xi z_h^{D-2}}\log\(\frac{\epsilon}{z_h}\) \nonumber \\
&& +\frac{V_\perp \ell^{2D-3}}{G_N \xi z_h^{D-2}}F(D) 
\eea
where $F(D)$ is undetermined function that depends only on the number of spacetime dimensions. Indeed one can evaluate the integral in a case by case basis, for example, for $D=4$ we obtain:
\bea
\CC^{\rm V}(L)&=&\frac{V_\perp \ell^{5}}{G_N \xi z_h^{2}} \int_{\delta/z_h}^{1-\epsilon/z_h} \frac{dx}{x^{3}\sqrt{1-4x^{3}+3x^{4}}} \nonumber \\
&=&\frac{V_\perp \ell^{5}}{2 G_N \xi }\(\frac{1}{\delta^2} +\sqrt{\frac{2}{3}}\frac{1}{z^2_h}\log\( 6(\sqrt{6}-2)\)-\sqrt{\frac{2}{3}}\frac{\log(\epsilon/{z_h})}{z_h^2}  \)
\eea
and for $D=5$
\bea
\CC^{\rm V}(L)&=&\frac{V_\perp \ell^{7}}{G_N \xi z_h^{3}} \int_{\delta/z_h}^{1-\epsilon/z_h} \frac{dx}{x^{4}\sqrt{1-3x^{4}+2x^{6}}} \nonumber \\
&=&\frac{V_\perp \ell^{5}}{3G_N \xi}\(\frac{1}{\delta^3} +\frac{\sqrt{3}}{2}\frac{\log(6)}{z^3_h}-\frac{\sqrt{3}}{2}\frac{\log({\epsilon}/{z_h})}{z_h^3}  \)\,.
\eea
The undetermined function $F(D)$ seems to have the structure
$\log(G(D))/\sqrt{2(D-1)(D-2)}$.

\paragraph{Weakly charged black holes:}
Another regime in which one can have analytic control is $q^2z_h^{2(D-2)}\ll 1$. In this case the horizon equation implies $1-mz_h^{D-1}+q^2z_h^{2(D-2)}=0$, which means that we can write $mz_h^{D-1}=1+q^2z_h^{2(D-2)}$ and do an expansion in powers of $q^2z_h^{2(D-2)}$. The integral of interest is
\bea \label{integral}
I&=&\int_{\delta/z_h}^{1} \frac{dx}{x^{D-1}\sqrt{1-m z_h^{D-1}x^{D-1}+q^2 z_h^{2(D-2)}x^{2(D-2)}} }\nonumber \\
&=& \int_{\delta/z_h}^{1} \frac{dx}{x^{D-1}\sqrt{1-x^{D-1}-q^2 z_h^{2(D-2)}x^{D-1}(1-x^{D-3})} }
\eea
where we have factored out the $z_h$ dependence up to the divergent UV piece and the dimensionless parameter $q^2 z_h^{2(D-2)}$.

Notice that one obvious concern in the previous equation is whether or not the perturbative expansion in $q^2 z_h^{2(D-2)}$ breaks down in the region where $x\approx 1$.
This is not the case, since the denominator of (\ref{integral}) behaves as $\sqrt{(1-x^{D-1})(1-\frac{D-3}{D-1}q^2 z_h^{2(D-2)})}$ in the $x\approx 1$ regime, and therefore the term proportional to $q^2$ is parametrically smaller for all $x$.

At first order in $q^2z_h^{2(D-2)}$ the integral is
\bea
I &=& \int_{\delta/z_h}^{1} \frac{dx}{x^{D-1}\sqrt{1-x^{D-1}} }+ \frac{q^2 z_h^{2(D-2)}}{2}\int_{\delta/z_h}^{1}dx \frac{1-x^{D-3}}{(1-x^{D-1})^{3/2} }
\nonumber \\
&\approx &\frac{(D-3)}{2(D-2)}\frac{\sqrt{\pi}\Gamma(\frac{D}{D-1})}{\Gamma(\frac{D+1}{2(D-1)})}  + \frac{1}{D-2} \frac{z^{D-2}_h}{\delta^{D-2}}  \nonumber \\
&& +\frac{q^2 z_h^{2(D-2)}}{D-1}\[\frac{(D-3)}{2}\frac{\sqrt{\pi}\Gamma(\frac{D}{D-1})}{\Gamma(\frac{D+1}{2(D-1)})}+
\frac{\sqrt{\pi}\Gamma(\frac{D-2}{D-1})}{\Gamma(\frac{D-3}{2(D-1)})} \]
\eea
and therefore the complexity of the weakly charged black hole is given by
\bea\label{qqq}
\CC^{\rm V}(L)=\frac{V_\perp \ell^{2D-3}}{G_N \xi}\Bigg[\frac{(D-3)}{2(D-2)}\frac{\sqrt{\pi}\Gamma(\frac{D}{D-1})}{\Gamma(\frac{D+1}{2(D-1)})}\(\frac{1}{z_h^{D-2}}+\frac{D-2}{D-1}q^2 z_h^{D-3} \) \nonumber \\
+ \frac{q^2 z_h^{D-3}}{(D-1)}\frac{\sqrt{\pi}\Gamma(\frac{D-2}{D-1})}{\Gamma(\frac{D-3}{2(D-1)})} + \frac{1}{(D-2)} \frac{1}{\delta^{D-2}} \Bigg]\,.
\eea
Notice that this answer is given in a mixed expansion, in which we used the exact $z_h$ and expanded the integral in powers of $q^2 z_h^2$, however a more consistent expansion would consider also $z_h=z_h^{(0)}+\delta z_h$ where $z_h^{(0)}$ is $z_h(q^2=0)$. It is interesting to note that in such expansion the expression simplifies to
\bea
\CC^{\rm V}(L)=\frac{V_\perp \ell^{2D-3}}{G_N \xi}\Bigg[\frac{(D-3)}{2(D-2)}\frac{\sqrt{\pi}\Gamma(\frac{D}{D-1})}{\Gamma(\frac{D+1}{2(D-1)})}\frac{1}{{z_h^{(0)}}^{D-2}} \nonumber \\
+ \frac{q^2 {z_h^{(0)}}^{D-3}}{(D-1)}\frac{\sqrt{\pi}\Gamma(\frac{D-2}{D-1})}{\Gamma(\frac{D-3}{2(D-1)})} + \frac{1}{(D-2)} \frac{1}{\delta^{D-2}} \Bigg]\,.
\eea
In this form we see explicitly that there is a legitimate $q^2$ correction to the zero charge case for the complexity which cannot be absorbed in the $q$ dependence of the new $z_h$ as observed from (\ref{qqq}).

\subsection{Shock wave}
\label{sec:shock}

An important motivation for the complexity equals action or complexity equals volume proposals were their linear growth behavior at late times \cite{CA, Roberts_LocalizedShocks,Stanford_ComplexityShocks}. This observation was more precisely stated in the CA duality by showing that for neutral black hole geometries one indeed has
\bea\label{late-timedC/dt}
\frac{d \CA}{dt_{L}}=2M\,,
\eea
at large $t_L$. In (\ref{late-timedC/dt}), the LHS is obtained by fixing the time slice on the right side of the boundary geometry $t_{R}=0$ and varying the left boundary time $t_{L}$. The complexification growth of the pure state is associated to the growth of the part of the action that lies behind the horizon. Since in the evaluation of subsystem complexity one only consider the region outside the horizon one does not expect a similar statement to hold in that case. In other words,  the action associated to the region $\CW_L \cap \mathcal{E}$ is invariant under $t_{L}\to t_{L} + \delta t$.

Nevertheless, one can consider a small deviation of the black hole geometry by perturbing the system slightly in the past. This is equivalent to injecting energy into the geometry in the form of a shock wave. In this case then the state is time dependent and its complexification rate is expected to have the same late time behavior, and what it is more it should present the expected time delay due to the injection of the shock wave \cite{CA}. In this section we evaluate the subsystem complexity in this dynamical situation for a neutral black hole with arbitrary asymptotic geometry. (To obtain the result for the flat boundary geometry we simply set $k$ to 0).

The metric we consider is
\bea
ds^{2}=-f(r)dt^{2}+\frac{dr^{2}}{f(r)}+r^{2} d \Sigma^{2}_{k, D-2}
\eea
where $d \Sigma^{2}_{k, D-2}$ is the line element of the boundary geometry which can be flat $(k=0)$, spherical $(k=1)$ or hyperbolic $(k=-1)$.

To describe the geometry in the presence of a shock wave it is convenient to move to light coordinates defined by
\bea
u&=& e^{\frac{2\pi}{\beta} (r^{*}(r)-t)} \nonumber \\
v&=& \mp e^{\frac{2\pi}{\beta} (r^{*}(r)+t)} \qquad
\eea
where the $-$ sign corresponds to the region $r>r_{h}$ while $+$ to $r<r_{h}$ and $r^{*}(r)$ is defined by $dr^{*}=dr/f(r)$. The parameter $\beta=4\pi/f'(r_{h})$.
Notice that  $r^{*}(r)$ decreases as $r$ decreases. If we fix $r^{*}(\infty)=0$ then $r^{*}$ is negative everywhere else (in these coordinates the boundary is located at $r\to \infty$). In particular, at the horizon $r^{*}(r_{h})=-\infty$, leading to $uv=0$.

 The back-reacted solution of this metric in the presence of a shock wave in null coordinates is given by
 \bea
ds^2=-2A(u,v)du dv+r^2d\Omega^2_{D-2}+2A(u,v)h \delta(u) du^2
\eea
where
\bea
A(u,v)=-\frac{2f(r)}{uv (f'(r_h))^2} \qquad \textrm {and}
\eea
\bea
f(r)=k-\frac{8\pi }{(D-2)\Sigma_{k,D-2}}\frac{2 G_N M}{r^{D-3}} +\frac{r^2}{\ell^2}\,.
\eea
Here $h\sim e^{\frac{2\pi}{\beta}(|t_w|-t_*)}$ is the shift produced by the shock wave, where $-|t_w|$ is the boundary time at which the operator dual to the shock wave is inserted. The back reacted energy created by the shock wave scales with $|t_w|$ as $E\sim e^{\frac{2\pi}{\beta}|t_w|}$ and $t_*=\frac{\beta}{2\pi} \log(\ell^{D-2}/G_N)$ is the scrambling time. Therefore, for $|t_w|<t_*$ the effect of the inserted energy has not effect on the geometry.

The net effect of the shock wave is to separate  the Penrose diagram along the $u=0$ region at which the shock wave has an important effect; however away from it the metric looks the same as the original black hole geometry. The essential difference will be in the way we glued the two sides together. Continuity of the $v$ coordinate then implies that one has to shift the two spacetimes along the $v$ coordinates by an amount $h$.

The horizon of the new metric will be located at $v=h$ as described by the original metric and therefore will be behind that horizon region. The subsystem complexity of the perturbed metric can then be computed using the unperturbed black hole metric but the region of interest includes both exterior and interior regions. One can split that region in two: one outside the black hole horizon and one behind the horizon using the additivity property of the action. Since we are interested in the time-dependent term in the action then we will only focus on the behind horizon region which as we mentioned gives the full time dependence.

\begin{figure}
$$
\includegraphics[scale=0.5]{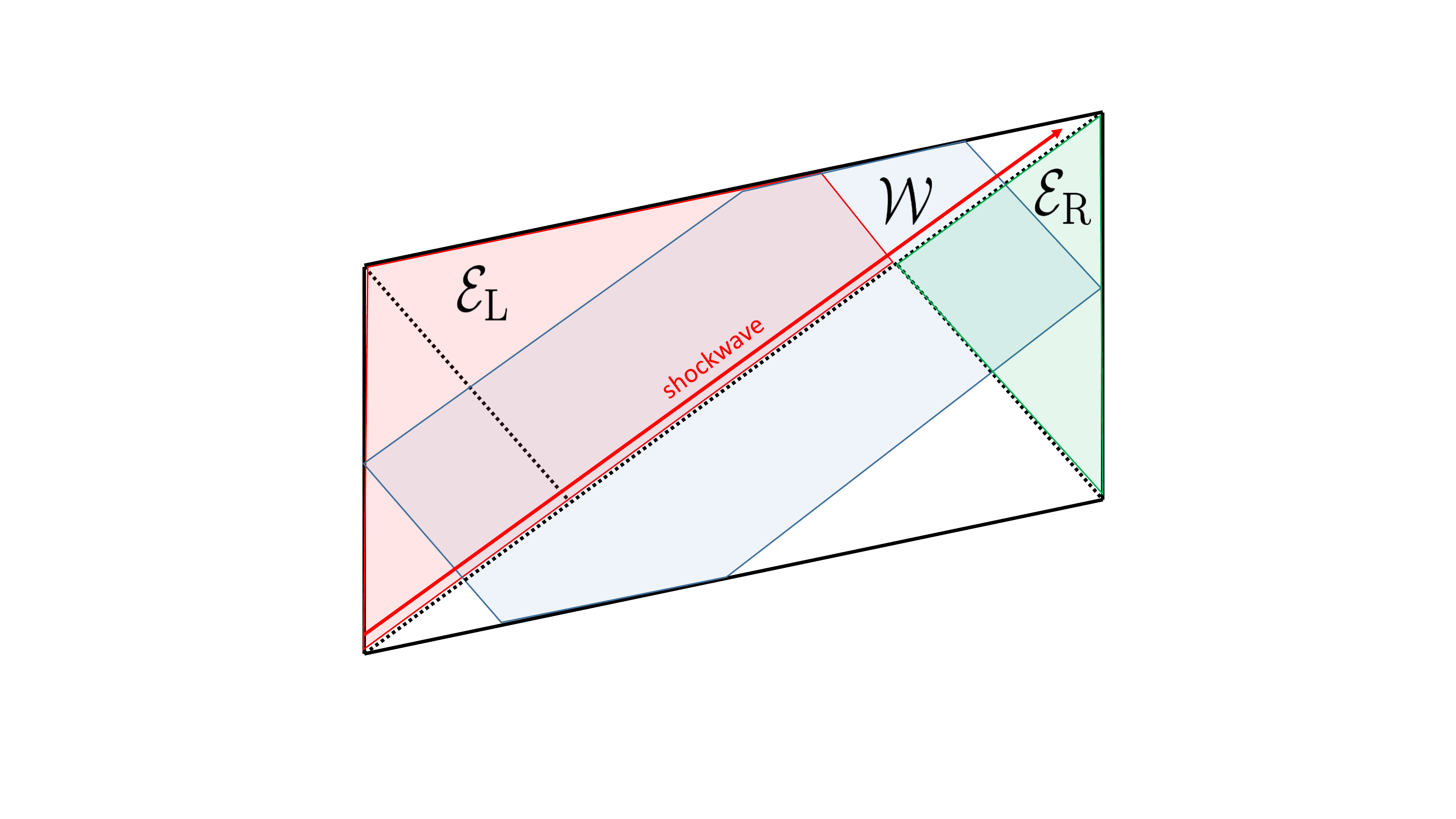}
 $$
\caption{ Representation of the left and right entanglement wedges $\CE_L$ and $\CE_R$ in red and green respectively, for an eternal black hole in the presence of a shock wave inserted on the left boundary. In blue it is represented the associated WdW patch}
\label{shockwave}
\end{figure}

The full action is given by
\bea\label{actionshock}
\CA_L&=&\frac{1}{16 \pi G_N}\int_{\CM}\sqrt{|g|}(R-2\Lambda)+\frac{1}{8\pi G_N}\int_{\partial \CM} \sqrt{|h|}K +\frac{1}{8 \pi G_N} \int_{(\partial \mathcal{M})_N} \Theta \log\(l_c |\Theta | \)dS  \nonumber \\
&& \qquad \qquad \qquad  \qquad \qquad \qquad+\sum_i \frac{1}{8\pi G_N}\int_{\Sigma'_i} a_i dS_i
\eea
associated to the region given by $\CE_L \cap \CW$ of figure \ref{shockwave}. The region $(\partial \mathcal{M})_N$ represents the boundary parts of the region $\CE_L \cap \CW$ which are null, and their respective integral represents the counter terms that renders the contributions to the null surfaces parametrization invariant. 

We would like to start with the evaluation of the first term, the bulk space-time integral. To perform this integral it is convenient to do a further change of coordinates $(\xi, \chi)$ given by $\xi=uv$ and $\chi=u/v$ on which the unperturbed metric looks

\bea
ds^2=-\frac{A}{2\xi}d\xi d\xi +\frac{A\xi}{2\chi^2}d\chi d\chi+r^2d \Omega^2_{D-2}
\eea
and the bulk integral is therefore given by
\bea\label{xichi}
\int \sqrt{|g|}d^Dx=\half \Sigma_{k,D-2}\int |A|r^{D-2}d\xi\frac{ d\chi}{\chi}
\eea
The bulk region is then delimited by the surfaces $\xi=1$ (the singularity), $\xi=\epsilon$ (the horizon when $\epsilon\to 0$), $v=h$ or equivalently $\chi=\xi/v^2=\xi/h^2$ and $\chi=u_0^2/\xi$. Both $A$ and $r$ are only functions of $\xi$ and therefore the $\chi$ integral can be performed directly
\bea
\int \sqrt{|g|}d^Dx &=&\half \Sigma_{k,D-2}\int_{\epsilon}^1 d\xi  |A|r^{D-2}\int_{\xi/h^2}^{u_0^2/\xi}\frac{ d\chi}{\chi} \nonumber \\
&=& -\frac{2\Sigma_{k,D-2}}{(f'(r_h))^2}\int_{\epsilon}^1  \frac{d\xi}{\xi}r^{D-2} f(r)(\log(u_0 h)-\log(\xi))
\eea
but from the definitions one has $d\xi/\xi=(4\pi/\beta)dr/f(r)$ then the $\xi$ integral turns into a trivial integral on $r$ where now the end points go from $r=r_h$ to $r=0$
\bea
\int \sqrt{|g|}d^Dx
&=& \Sigma_{k,D-2}\(\frac{\beta}{2\pi}\)\int^{r_h}_0 d rr^{D-2} (\log(u_0 h)-\log(\xi))
\eea
We are interested in the term proportional to the initial conditions and shock wave parameters, therefore we can ignore the $\log(\xi)$ term. In that case we have
\bea
\frac{1}{16\pi G_N}\int \sqrt{|g|}(R-2\Lambda)d^Dx
&=&- \frac{\Sigma_{k,D-2}r_h^{D-1}}{8\pi G_N \ell^2}\(\frac{\beta}{2\pi}\)\log(u_0 h)
\eea
Using $u_{0}=e^{\frac{2\pi}{\beta} t_{L}}$ and $h\sim e^{\frac{2\pi}{\beta}(|t_{\omega}|-t_{*})}$ this contribution turns into
\bea
\frac{1}{16\pi G_N}\int \sqrt{|g|}(R-2\Lambda)d^Dx
&=&- \frac{\Sigma_{k,D-2}r_h^{D-1}}{8\pi G_N l^2}(t_{L}+|t_{\omega}|-t_{*})\,.
\eea

The second term in (\ref{actionshock}) leads to three boundary null surface terms which give zero contribution to the action, for affinely parametrized null generators, and one timelike boundary surface surrounding the singularity which we will compute as follows. First, the timelike boundary surface $\partial \CM$ is given by $\xi=1$. Here $K=\half n^\alpha \partial_\alpha \log{|h|}$ where the line element associated to $h$ is simply
\bea
ds^2|_{\xi=cons}=\frac{A\xi}{2\chi^2}d\chi d\chi+r^2 d\Sigma^2_{k,D-2}
\eea
and $n^\alpha=\delta^{\alpha}_\xi/\sqrt{g_{\xi \xi}}$, $|h|=\frac{-f(r)\,r^{2(D-2)}}{\chi^2(f'(r_h))^2}$ and $g_{\xi \xi }=\frac{-f(r)}{\xi^2(f'(r_h))^2}$\,. Simplifying the expressions we get
\bea
\int_{\partial \CM}\sqrt{|h|}K=\Sigma_{k,D-2}r^{D-2}\(-\frac{\beta}{4\pi}\)\(\frac{2(D-2)f(r)}{r}+f'(r)\)(\log(u_0 h)-\log(\xi)) \nonumber \\
\eea
where the $\chi$ integral was carried out from $\chi=\xi/h^2$ to $\chi=u_0^2/\xi$ as in the bulk case.
Here, $\xi=1$ cancels the term proportional to $\log \xi$ which we ignored for the bulk contribution. Multiplying by $1/8\pi G_N$ and evaluating the previous result on $r=0$ gives us the boundary contribution to the complexity which is
\bea
\frac{1}{8\pi G_N}\int_{\partial \CM}\sqrt{|h|}K&=&\( \frac{D-1}{D-2}\)M \(\frac{\beta}{2\pi}\)\log(u_0 h)\nonumber \\
&=&\( \frac{D-1}{D-2}\)M(t_{L}+|t_{\omega}|-t_{*})\,.
\eea
The calculation of the third term in (\ref{actionshock}) goes exactly as in (\ref{eternalBH}), in other words, the null surfaces at the black hole horizons do not contribute and the null surfaces that hit the future singularity do contribute but it does it so that the full contribution is time dependent. The reason is that one can parametrize such integrals with the affine parameter $\lambda=-\ell/z$ which runs from $z_h$ to $\infty$ and it is independent of the enlargement of the black hole interior due to shock wave.

Finally we will focus on evaluating the terms coming from the corner contributions obtained at the intersection of light-like surfaces as the ones appearing on the horizon. The corners appearing on the singularity surface do not contribute as the volume factor goes to zero as $r\to 0$. The calculation goes in the exact same way as in the evaluation of the corner contributions of the subsystem complexity for neutral black holes of section (\ref{eternalBH}). The only difference here is that the regularized corners lie behind the horizon region and the individual contributions have a slightly different form in  the $(t,r)$ coordinates, which is
\bea
\int_{\Sigma'_i} a\, dS=-\textrm{sgn}_i \, \Sigma_{k,D-2}r^{D-2}\log\(-\frac{f(r)}{c \bc}\) \,.
\eea

Here, as described in Appendix \ref{AA}, we add and subtract the corner term appearing at the intersection $H^+\cap H^-$ and rewrite the differences of the corners in terms of the logarithms of $uv$ products, leading to

\bea
\frac{1}{8\pi G_N}\sum_i \int_{\Sigma_i} a_i dS_i=\frac{\Sigma_{k,D-2}r_h^{D-2}}{8\pi G_N}\left[ \log\(u_0\, h\) -\log\(\eu\,\ev\)+\log\(\frac{-f(r_{\eu\, \ev})}{c\bc}\)\right] \nonumber  \\
\eea
where $r_{\eu\, \ev}$ is the regularized radial coordinate lying at the intersection of the light sheet $u=\eu$ and $v=\ev$. The last two terms in the $\eu, \ev\to 0$ limit gives a finite contribution which is independent of the parameters $u_0$, $h$ and therefore we ignore them here. Notice however that these contributions could lead to divergent terms as the ones we found in previous calculations but those terms would not be time dependent. The time dependent piece is given by
\bea
\frac{1}{8\pi G_N}\sum_i \int_{\Sigma_i} a_i dS_i=\frac{\Sigma_{k,D-2}r_h^{D-2}}{16\pi G_N}f'(r_{h})(t_{L}+|t_{\omega}|-t_{*})
\eea
where we have used the relations $u_{0}=e^{\frac{2\pi}{\beta} t_{L}}$ and $h\sim e^{\frac{2\pi}{\beta}(|t_{\omega}|-t_{*})}$.

Adding up all the pieces one gets
\bea
\CC^{\rm A}(L)=2M(t_{L}+|t_{\omega}|-t_{*})
\eea
and therefore at large $t_L$ we reproduced the late time complexification rate for subsystem complexity with the proper time shift due to the switchback effect \cite{Stanford_ComplexityShocks, CA}. Indeed, this effect provided important evidence for the CA and CV conjectures, which in the context of pure state complexity was tested even in the presence of multiple shock waves \cite{Shenker_MultiShocks, CA}. The result for the complexification rate is then
\bea
\frac{d\CC^{\rm A}(L)}{d t_L}=2M
\eea
for large $t_L$.

\section{Measures of mixed-state complexity}
\label{sec:measures}

In the previous section, we calculated the volume and action quantities $\CC^{\rm V}$ and $\CC^{\rm A}$ for thermal states of holographic systems. In the spirit of the CV and CA conjectures, we would like to relate these to some notion of complexity for mixed states. Therefore, our first task is to come up with measures of complexity for mixed states. We will find that there are many ways to define such measures, and it is not straightforward to determine the relations among them. This is perhaps not surprising, as a similar situation obtains for entanglement in bipartite mixed states; many different measures have been defined (entanglement of purification, entanglement of formation, entanglement of distillation, logarithmic negativity, and so on), and determining how they are related to each other is far from straightforward.

In subsection \ref{sec:defs}, after reminding ourselves of the relevant notion of complexity for pure states, we define several measures of complexity for mixed states. In subsection \ref{sec:expectations}, we estimate the values of these measures in thermal states, in particular their dependence on temperature, using intuition from tensor networks. Then in subsection \ref{sec:comparison} we compare these estimates to the values of $\CC^{\rm V}$ and $\CC^{\rm A}$ obtained in section \ref{sec:holographic}. We find that one of our proposed definitions matches well (to within the precision of our estimates) the behavior of $\CC^{\rm A}$. We thus arrive at a concrete and well-motivated subsystem CA conjecture. On the other hand, we do not find a match between $\CC^{\rm V}$ and any of our proposed complexity definitions. In subsection \ref{sec:otherdefs}, we briefly explore other possible approaches to defining mixed-state complexity, but again fail to find a plausible match to $\CC^{\rm V}$.

It is worth reiterating that almost all of the mixed states we consider in this paper are thermal (i.e.\ Gibbs or generalized Gibbs) states. This is both an advantage, as it gives us a handle on estimating their complexities that we would not necessarily have for general states, and a limitation. In particular, these states are static, eliminating the whole issue of time dependence, which was central to the development of the original CV and CA conjectures \cite{Susskind_Complexity, CA}. To further test and explore our subsystem CA conjecture, it will be important to study other types of subsystems, in particular those in time-dependent states. We took a small step in this direction in subsection \ref{sec:shock} where we studied subsystem complexity for a time-dependent shockwave geometry.

\subsection{Proposed definitions}\label{sec:defs}

We begin with simplest notion of pure state complexity. This definition has three ingredients: a reference state, a set of allowed gates, and a tolerance. The complexity of a target pure state is defined as the minimum number of gates from the allowed set needed to take the reference state to the target state up to the specified tolerance. The notion of tolerance has considerable freedom in it. We could require that the target state and the evolved reference state are close in trace norm or we could demand that they have approximately equal expectation values for some operators or any of a myriad of other measures. Let us denote this measure of pure state complexity, for some fixed set of choices, by $\mathcal{C}$. We note that some pure state schemes which are particularly adapted to the problem of field theory complexity have been explored recently~\cite{Nielsen_ComplexityGeometry,Jefferson_FreeQFTComplexity,Chapman_FreeQFTComplexity}.

To approach the problem of mixed state complexity, we begin by making some preliminary remarks. First, note that the definition of $\mathcal{C}$ made no reference to ancilla, meaning that we implicitly fixed the number of qubits and only allowed gates to act on those qubits. However, one could also consider notions of complexity with ancilla included. We could either allow no ancilla, allow ancilla but require them to return approximately to their initial state, or allow ancilla with arbitrary final states so long as the target state is approximately obtained. These definitions are not all equivalent, although it is not clear under what conditions they differ substantially. We will assume, as above, that the definition of pure-state complexity does not allow ancilla even at intermediate stages.

Second, observe that there is a potential distinction between mixed states and subsystem states (which may of course still be mixed). A complexity measure for mixed states must be applicable to any mixed state without reference to any other system. However, a complexity measure for subsystem states could depend on the state of the whole system as well. It is not obvious which notion is relevant for holography, but we proceed by thinking about mixed states without reference to a fixed purification.

Third, we will demand that our notion of mixed state complexity reduce to the pure state definition when the state is pure. This seems trivial, but it turns out to restrict the kinds of operations we can consider, e.g., we cannot allow ancilla in the mixed case unless we also allow them in the pure case.

With the above issues in mind, we now present two approaches to defining mixed-state complexity. Our analysis is complementary to some discussions in the quantum information literature~\cite{1998quant.ph..6029A}.

\textbf{Purification approach:} The simplest definition of mixed state complexity is phrased in terms of minimal purifications. Given a mixed state $\rho$ on $n$ qubits, an initial state $| 0\ldots0\rangle$, a set of allowed unitary transformations $G$, and a tolerance $\epsilon$, the \emph{purification complexity} $\mathcal{C}_P$ of $\rho$ is defined as the minimum number of gates from $G$ needed to transform the initial pure state plus an arbitrary number of ancilla qubits initialized into the state $|0\rangle$ into a purification of $\rho$ up to tolerance $\epsilon$. Ancilla may only be used if they are entangled with the $n$ qubit system at the end of the process. This is an important restriction if we are to recover the ancilla-less definition of pure state complexity (to recover a pure target state, all ancilla must be unentangled with the system up to the tolerance). Roughly speaking, this definition may be summarized as the pure state complexity of the minimum complexity purification of $\rho$ where we use only essential ancilla.

A few additional comments are in order to clarify our hypothesis for the nature of the essential ancilla. If the goal is the minimize the use of ancilla, the first step would be to restrict the system plus purifier to be defined on no more than double the number of degrees of freedom of the original system. Moreover, when the system state is pure, then no ancilla should be used. Since the basis of the purifier state has no restriction placed on it, it is natural to suppose that the number of ancilla qubits is proportional to the entropy of the system state. For example, there is no need to embed the order $e^S$ states of the purifier that are entangled with the system into the larger UV Hilbert space for the purifier.

\textbf{Spectrum approach:} Another way to think about complexity for a mixed state is to break the problem of creating the state into two parts: creating its spectrum and creating its basis of eigenstates. Given a mixed state $\rho$, an initial state $| 0\ldots0\rangle$, a set of allowed unitary transformations $G$, and a tolerance $\epsilon$, we define the \emph{spectrum complexity} $\mathcal{C}_S$ of $\rho$ as the minimum number of unitaries from $G$ needed to transform the initial state plus ancilla into a state whose partial trace has the same spectrum as $\rho$ and such that all ancilla are entangled with the original system. Since $\rho$ has the same spectrum as itself, in general $\mathcal{C}_S \leq \mathcal{C}_P$.

Defining the complexity to construct the basis of eigenstates is harder. We could try to define it as the minimum number of unitaries needed to transform the initial state plus ancilla into a state whose partial trace has the same basis as $\rho$. However, since the maximally mixed state has the same basis as any state $\rho$, it would follow that the complexity to construct the basis of any state $\rho$ is upper bounded by a fixed number independent of $\rho$ of order the number of qubits.

We will therefore suggest two other definitions of basis complexity. First, since $\mathcal{C}_S \leq \mathcal{C}_P$, we could define the basis complexity as their difference:
\begin{equation}
\mathcal{C}_B := \mathcal{C}_P - \mathcal{C}_S \geq 0\,,
\end{equation}
i.e.\ roughly the extra work needed to get the basis right. Note that it is not really clear whether the effort is exactly additive in this fashion, e.g.\ it might be roughly as hard to prepare just the spectrum as to prepare the whole state.

Alternatively, we could define the basis complexity by starting with the minimal complexity state $\rho_{\text{spec}}$ with the same spectrum as $\rho$, and then finding the minimum number of gates needed to change $\rho_{\text{spec}}$ into $\rho$. This is always possible precisely because $\rho$ and $\rho_{\text{spec}}$ share the same spectrum. We denote this notion of basis complexity by $\tilde{\mathcal{C}}_B$.

As usual, it is not clear how $\mathcal{C}$ and $\tilde{\mathcal{C}}_B$ are related in general, but since $\mathcal{C}_P \leq \mathcal{C}_S + \tilde{\mathcal{C}}_B$ (because reaching $\rho$ via $\rho_{\text{spec}}$ is one possible circuit) it follows that
\begin{equation}
\tilde{\mathcal{C}}_B \geq \mathcal{C}_B.
\end{equation}

\begin{figure}
  \centering
  \includegraphics[width=0.5\textwidth]{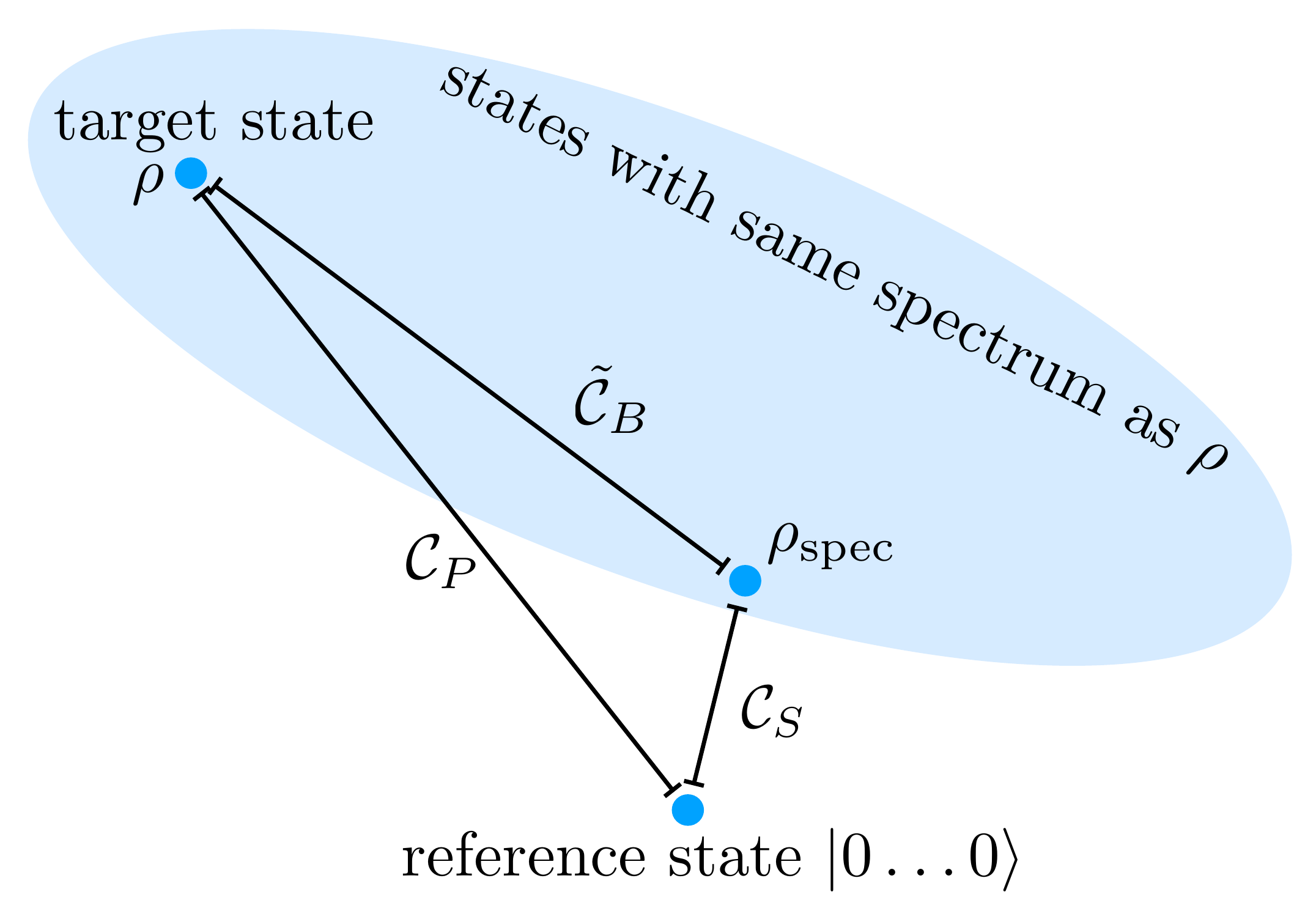}
\caption{Illustration of measures of complexity defined in the main text. Roughly speaking, $\CC_P$ is the minimum number of gates required to go from the reference state to the target state $\rho$. Among the states with the same spectrum as $\rho$ (blue region), the one that can be obtained with the fewest gates starting from the reference state is called $\rho_{\rm spec}$, and the minimum number of gates is $\CC_S$. $\tilde\CC_B$ is the minimum number of gates required to take $\rho_{\rm spec}$ to $\rho$. $\CC_B$ (not shown) is $\CC_P-\CC_S$, and by the triangle inequality this cannot be more than $\tilde\CC_B$. (More precisely, to go from the reference state to some mixed state such as $\rho$ or $\rho_{\rm spec}$, we first add an arbitrary number of ancilla qubits to the reference state and then act with the gates to obtain a purification of the mixed state, in which all ancilla are required to be entangled with the original system.)
}\label{fig:complexitydefs}
\end{figure}

For a pure state of complexity $\CC$, it is easy to see that $\CC_S=0$ while $\CC_B=\tilde\CC_B=\CC$. Thus, in some sense, the basis complexity (with either definition) is the analogue of pure-state complexity, while the spectrum complexity is a new feature of mixed states. These various definitions are illustrated in figure \ref{fig:complexitydefs}.

\subsection{Expectations from tensor networks}\label{sec:expectations}

To give a sense of these definitions and how they behave in a field-theory context, let us imagine applying them to a chaotic spin chain whose low-energy physics is described by a strongly interacting conformal field theory which has central charge larger than one and is chaotic. Below, these expectations will be compared with the results of holographic calculations. To fix notation, suppose that the model consists of $n$ qubits with Hamiltonian $H$. The Hamiltonian has energies $E_i$ and eigenvectors $|i\rangle$. Throughout we consider the thermal state, $\rho \propto e^{-H/T}$. We focus on the two extremes: $T=0$ and $T=\infty$.

At zero temperature, the ground state has approximate conformal invariance. Assuming it has a renormalization group circuit which prepares the ground state, e.g., a MERA-like circuit, the pure-state complexity of the ground state is of order $n$, say $\mathcal{C}(T=0) = k_1 n$. Since the state is pure, it has trivial spectrum and we find
\begin{equation}
T=0:\qquad
\mathcal{C}_P = k_1 n\,,\qquad
\mathcal{C}_S=0\,,\qquad
\mathcal{C}_B = \tilde{\mathcal{C}}_B = k_1 n\,.
\end{equation}

At infinite temperature, the thermal state is a maximally mixed state. In this case one finds
\begin{equation}
T=\infty:\qquad
\mathcal{C}_P= \mathcal{C}_S = k_2 n\,,\qquad
\mathcal{C}_B = \tilde{\mathcal{C}}_B = 0\,.
\end{equation}
These statements follow because any state with the right (uniform) spectrum is automatically the right state, and because the maximally mixed state can be prepared with order $n$ gates using $n$ ancilla.

Based on these two limits, we can make a minimal guess for the temperature dependence of the various complexity measures. We observe that $\mathcal{C}_P$ need not depend strongly on temperature, although of course it could have some temperature dependence. Meanwhile, $\mathcal{C}_S$ should increase while $\mathcal{C}_B$ and $\tilde{\mathcal{C}}_B$ should decrease as a function of temperature, although again we obviously cannot rule out non-monotonicity. Furthermore, it seems reasonable to suppose that $\mathcal{C}_B$ and $\tilde{\mathcal{C}}_B$ are of the same order for all temperatures.

We can use intuition from tensor networks to be a bit more specific about the form of these complexities at intermediate temperatures. If we imagine that the minimal circuit which prepares a purification of the thermal state has two pieces, one which prepares the spectrum and one which prepares the basis, then it is natural to guess that
\begin{equation}
\mathcal{C}_S \sim \alpha S
\end{equation}
where $S$ is the thermal entropy and
\begin{equation}
\mathcal{C}_B \sim \tilde{\mathcal{C}}_B \sim k_1 n - \beta S\,.
\end{equation}
In a MERA-like circuit, these behaviors come from two distinct effects: (1) The spectrum must be prepared, and if the spectrum may in some sense be approximated as Bell pairs, then the complexity should be roughly proportional to the entropy. (2) The basis must be prepared, but whereas the ground state has long-range correlations, the mixed state has shorter-range correlations, so less of the renormalization group part of the circuit is needed. Counting gates shows that the reduction is also roughly proportional to the entropy. However, it is not clear how the coefficients $\alpha$ and $\beta$ compare or how they depend on temperature, e.g., due to logarithmic factors. Hence it is not clear at this level how $\mathcal{C}_P = \mathcal{C}_S + \mathcal{C}_B \sim k_1 n + (\alpha-\beta)S$ depends on temperature.\footnote{It's worth noting that we have good reason to believe the TFD state is not the minimal purification. Specifically, to use the basis and spectrum language, the TFD has many ``useless'' gates that adjust the basis of the purifying system. This basis change does not effect the state of the original system, hence it is non-minimal.}

\subsection{Comparison to holographic calculations}\label{sec:comparison}

We now compare our various definitions and expectations to the holographic computations described above. For simplicity, we focus on the uncharged eternal black hole in any dimension. In the thermofield double state we found that CA generically predicted either
(see \eqref{fullpart})
\begin{equation}
2 \mathcal{C}^{\rm A}(\rho) > \mathcal{C}^{\rm A}(|\text{TFD}\rangle) \,\,\,\, \text{(subadditive)}
\end{equation}
or
\begin{equation}
2 \mathcal{C}^{\rm A}(\rho) < \mathcal{C}^{\rm A}(|\text{TFD}\rangle) \,\,\,\, \text{(superadditive)}
\end{equation}
while CV predicted (see \eqref{fullpartV})
\begin{equation}
2\mathcal{C}^{\rm V}(\rho) = \mathcal{C}^{\rm V}(|\text{TFD}\rangle)\,.
\end{equation}
The relation between these quantities and the entropy was
\begin{equation}
\mathcal{C}^{\rm A}(\rho) \sim n \pm S
\end{equation}
(see \eqref{complexT}) and
\begin{equation}
\mathcal{C}^{\rm V}(\rho)\sim n + S.
\end{equation}
(see \eqref{volume-complexity}). Here $n$ stands for $V_\perp/\delta^{D-2}$ (the volume of the CFT in cutoff units), we have dropped logarithmic divergences, and we only care about the sign of the coefficient in front of the $S$ terms (in $D=3$ the coefficient of $S$ in the volume is exactly zero).

Are any of the quantities $\mathcal{C}_P$, $\mathcal{C}_S$, $\mathcal{C}_B$, and $\tilde{\mathcal{C}}_B$ consistent with CA? Suppose first that $\mathcal{C}^{\rm A}$ increases with temperature and is subadditive. Then the temperature dependence rules out interpreting it as $\mathcal{C}_B$ or $\tilde\CC_B$ because we expect the latter to decrease with temperature. Under the plausible assumption that the spectrum can be prepared without preparing the whole UV of the field theory, it follows that $\mathcal{C}^{\rm A}$ can also not be interpreted as $\mathcal{C}_S$ since the former is UV sensitive. Moreover, $\mathcal{C}_S$ does not reduce to the pure state definition of complexity. On the other hand, $\mathcal{C}_P$ does appear consistent with our expectations and the CA results. In particular, if we think of $\mathcal{C}_P$ as roughly the cost of the spectrum plus the cost of the basis, then because we must prepare the spectrum twice when preparing two copies of $\rho$ but only once when preparing $|\text{TFD}\rangle$, it follows that $2\mathcal{C}_P(\rho) > \mathcal{C}(|\text{TFD}\rangle)$ which corresponds to a subadditive $\mathcal{C}^{\rm A}$.

Now suppose that $\mathcal{C}^{\rm A}$ decreases with temperature and is superadditive. Again because of the UV divergence, the spectrum complexity is not a good match. The purification complexity is also no longer a good match since it should be subadditive. However, the basis complexity is now a better match. In particular, the basis complexity should decrease with temperature and be superadditive. This is because the TFD complexity should be roughly the spectrum complexity plus twice the basis complexity of a single side (for the left and right sides), hence it should be greater than twice the basis complexity of a single side.

What about CV? For similar reasons to the superadditive CA case, $\mathcal{C}_S$, $\mathcal{C}_B$, and $\tilde\CC_B$ are ruled out. Interestingly, $\mathcal{C}_P$ is also ruled out since we have $2\mathcal{C}^{\rm V}(\rho) = \mathcal{C}^{\rm V}(|\text{TFD}\rangle)$. Equality here is inconsistent with our previous story about basis plus spectrum unless the cost of the spectrum is zero.

A similar analysis can be applied to the case of charged black holes. For weakly or moderately charged black holes, we find that, within the precision of our analysis, $\mathcal{C}^{\rm A}$ can be qualitatively matched to the purification complexity or the basis complexity. The extremal limit is an interesting further probe of complexity/geometry duality, since we find that both measures diverge there. It will be interesting to explore the possible physical significance of this divergence in the future, since it seems unexpected from the point of view of boundary complexity.

\subsection{Other definitions}\label{sec:otherdefs}

The conclusion of the preceding analysis is that CA duality appears consistent with the $\mathcal{C}_P$ definition of mixed state complexity. By contrast, CV duality cannot apparently be consistently interpreted in terms of $\mathcal{C}_P$ unless our analysis in terms spectrum and basis is very misguided. However, this analysis is corroborated in its broad outlines by a tensor network picture of the thermal state. Confronted with these conclusions, we now expand the discussion to include other possible definitions of complexity.

\textbf{Open system approach:} Given a mixed state $\rho$, an initial state $\rho_0= |0 ...0\rangle \langle 0 ... 0|$, a set of allowed quantum operations\footnote{Formally, by quantum operation we mean a completely positive trace-preserving map.} $G$, and a tolerance $\epsilon$, the \emph{open system complexity} $\mathcal{C}_O$ of $\rho$ is defined as the minimum of number of operations from $G$ needed to transform the initial state into $\rho$ up to tolerance, $\epsilon$ say in trace norm. Formally, the open system complexity $\mathcal{C}_O$ is the minimum number such that
\begin{equation}
\rho \approx_{\epsilon} \Phi_{\mathcal{C}_O}\circ\cdots\circ \Phi_1(\rho_0)
\end{equation}
where $\Phi_i \in G$. We could obviously modify this definition by weighting elements in $G$ differently, by adjusting how the tolerance is defined, or by changing the initial state. Note that even if $\rho$ is a pure state, it is possible that by allowing general quantum operations, as opposed to just unitary transformations, some states could be reached more quickly.

Since allowing more general quantum operations, i.e., unitaries acting also on ancilla, does not reduce to the ancilla-less definition of pure state complexity, it follows that $\mathcal{C}_O$ can give different results than $\mathcal{C}$ when applied to pure states. It is not clear to us if the results can be vastly different, but we do know of cases where there is some difference. For example, in the context of quantum many-body physics, it is known that a Chern insulator ground state cannot be prepared by a finite depth circuit without ancilla, but two copies of a Chern insulator ground state (really a copy and a conjugate copy) can be prepared by a finite depth circuit. Here the inclusion or not of ancilla makes a substantial difference.

From the perspective of holography and tensor networks, it seems to us that ancilla have generally not played a role in the discussion. In other words, the general point of view has been that complexity should be defined with respect to the intrinsic resources of the system and should not make reference to auxiliary degrees of freedom. From this point of view, it makes more sense to think of the purifying system as physical, i.e., as instantiated in the rest of the geometry, instead of merely as ancilla used to apply more general quantum operations to a state. One concrete difference between the two points of view is in terms of whether or not we can act repeatedly on the ancilla.

\textbf{Ensemble approach:} An alternative point of view on mixed-state complexity arises from the fact that a mixed state can be written as a convex combination (i.e.\ ensemble) of pure states:
\begin{equation}
\rho = \sum_i p_i |\phi_i \rangle \langle \phi_i |\,.
\end{equation}
We can thus define the \emph{ensemble complexity} of $\rho$ as the corresponding convex combination of complexities of the elements $|\phi_i\rangle$, minimized over ways of writing $\rho$ as an ensemble:
\begin{equation}
\mathcal{C}_E(\rho) = \min_{\text{ensembles}\atop\{p_i,|\phi_i\rangle\}} \sum_i p_i\, \mathcal{C}(|\phi_i \rangle)\,.
\end{equation}
Note that the eigenbasis of $\rho$ is only one possible ensemble, and may be far from the minimal one. Furthermore, the states in a given ensemble need not be orthonormal, e.g., they could be overcomplete.

This ensemble-based definition seems qualitatively different from the other definitions given above, although we can relate it to them in some cases. It does have the virtue of reducing to the pure state complexity when the state $\rho$ is pure. One reason for considering this notion of complexity is that none of the other options we considered seemed to be a very good match for $\mathcal{C}^{\rm V}$. As we will explain below, however, $\mathcal{C}_E$ is roughly consistent with $\mathcal{C}^{\rm A}$, but it also does not seem very well suited to $\mathcal{C}^{\rm V}$.

What are our expectations for $\mathcal{C}_E$ within the spin chain model considered above? At zero temperature it should agree with $\mathcal{C}_P$ which is just the pure state complexity of the ground state. At infinite temperature the minimal complexity ensemble is simply the ensemble of product states. Hence we have
\beq
T=0:\qquad\mathcal{C}_E = k_1 n
\eeq
and
\beq
T=\infty:\qquad\mathcal{C}_E \propto n\,.
\eeq

If we tried to match these expectations to CV duality, we would be faced with the unusual conclusion that the ensemble complexity of the thermal state is always exactly equal to the complexity of the thermofield double state for all temperatures. While we are not aware of anything ruling this out, this seems unlikely to be true. For example, we can definitely find models, e.g., models with a trivial tensor product ground state, in which $\mathcal{C}_E$ is strongly dependent on temperature.

\subsection{Bounds on subsystem complexity}

The ensemble approach seems to have certain advantage over the other definitions, as it seems more tractable to explicit evaluation. To illustrate this point we compute a bound on the ensemble complexity (relative to the ground state) following the work of \cite{Yang:2017nfn}. For some work towards defining complexity in quantum field theory see \cite{Jefferson:2017sdb,Chapman:2017rqy,Khan:2018rzm,Yang:2018nda}. In \cite{Yang:2017nfn} the authors argued that the relative pure state complexity (which refers to the minimun number of gates required to take the vacuum state to any other pure state) associated to the coherent state $|re^{i\theta}\rangle=e^{-r^2}e^{r e^{i\theta}a^\dagger}|0\rangle$ is given by
\bea
\CC (|re^{i\theta}\rangle,|0\rangle)=r (|\cos\theta |+|\theta||\sin \theta|)\,,
\eea
where $a^\dagger$ is the creation operator associated to a single simple harmonic oscillator system, and $\theta$ is an angular coordinate with range in $[-\pi , \pi)$.

We would like to make a slightly weaker assumption and propose the formula
\bea \label{relcocom}
\CC (|re^{i\theta}\rangle,|0\rangle)=r f(\theta)
\eea
where $f(\theta)$ is undetermined, and use this simple result as a way to illustrate a simple way of finding useful bounds to the ensemble complexity. This requires knowledge of a formula for the pure state complexity as a well as a good candidate of low complexity ensemble.

We would like to illustrate this in the simplest example of a single harmonic oscillator system as well as its generalization to free quantum field scalar theory.

\subsubsection{Single oscillator}

Consider a single oscillator mode of a quantum mechanical system with Hamiltonian
\beq
H = \omega_0 a^\dagger a
\eeq
where $[a,a^\dagger]=1$. We would like to bound the ensemble complexity associated to a thermal state $\rho_\beta\equiv e^{-\beta H}$, where $\beta=1/T$ is the inverse temperature.

The thermal state, which one would normally write in the Hamiltonian basis $\{|n\rangle \}$ as
\bea
\rho_\beta=\frac{1}{Z}\sum_n e^{-\beta E_n}|n \rangle \langle n|
\eea
with $Z=1/(1-e^{-\beta \omega_0})$, has an equivalent decomposition in terms of the normalized coherent states $| r e^{i \theta} \rangle$ which are obtained from the vacuum by local unitary transformations and therefore are of relatively low complexity. Indeed, it is easy to check that $\rho_\beta$ is also given by
\bea\label{thermalcoherent}
\rho_\beta=\frac{A}{\pi} \int d\theta dr r e^{- A r^2} | r e^{i \theta} \rangle \langle r e^{i \theta} |
\eea
with $A = e^{\beta \omega_0}-1$, using the relation
\beq
| r e^{i \theta} \rangle = e^{-r^2/2}\sum_{n=0}^\infty \frac{r^n e^{i n \theta}}{\sqrt{n!}} | n\rangle \,.
\eeq

This ensemble then represents a relatively low complexity ensemble whose complexity can be used to bound our ensemble complexity $\CC_E$. That is, given
\bea
\CC_E=\text{min}_{\text{ensemble}}\sum_i p_i \CC(|\phi_i\rangle)
\eea
therefore
\bea \label{bound}
\CC_E\leq \frac{A}{\pi} \int d\theta dr\, r e^{- A r^2} \CC(| r e^{i\theta }\rangle)=\CC^{b}_E\,.
\eea
We would like to estimate the value of the upper bound on $\CC_E$,
$\CC_E^{b}$, using the formula for the pure state complexity of coherent states given by (\ref{relcocom}). However, we don't have a formula for $\CC(| r e^{i\theta }\rangle)$ but instead for $\CC(| r e^{i\theta }\rangle,|0 \rangle)$, therefore we obtain a bound on $\Delta \CC_E$ defined as\footnote{A similar quantity called complexity of formation was central in the discussion of \cite{Chapman:2016hwi}.}
\bea
\Delta\CC_E=\text{min}_{\text{ensemble}}\sum_i p_i \CC(|\phi_i\rangle,|0\rangle)
\eea
and which we denote as $\Delta\CC_E^{b}$. This is
\bea
\Delta \CC^{b}_E &=& \frac{A \(\int d\theta f(\theta)\)}{\pi} \int dr r^2 e^{ Ar^2}=\frac{\(\int d\theta f(\theta)\)}{4\sqrt{\pi A}}\nonumber \\
&=& \frac{\(\int d\theta f(\theta)\)}{4\sqrt{\pi}}\(\frac{e^{-\beta \omega_0}}{1-e^{-\beta \omega_0}}\)^{1/2}\sim \langle n \rangle^{1/2}\,.
\eea

One can try relate this answer with the thermodynamic quantities for the single oscillator
\bea
Z=\frac{1}{1-e^{-\beta \omega_0}}, \qquad \langle n \rangle=\frac{e^{-\beta \omega_0}}{1-e^{-\beta \omega_0}}
\eea
and
\bea
S=-\log(1-e^{-\beta \omega_0})+\beta \omega_0 \langle n \rangle\,.
\eea

This is more naturally achieved in the two extreme cases of low and high temperatures: For low temperature  $ \Delta \CC_E^b \sim e^{{-\beta \omega_{0}/2 }}$ and $S \sim (\beta \omega_{0})e^{{-\beta \omega_{0} }}$ therefore
\bea
\Delta \CC_E^b \sim S^{1/2}/(\log(S))^{1/2}\,.
\eea
Similarly for high temperature $\Delta\CC_E^b \sim 1/(\beta \omega_{0})^{1/2}$, while $S\sim -\log(\beta \omega_{0})$ and therefore
\bea
\Delta \CC_E^b \sim e^{S/2}.
\eea

\subsubsection{Free scalar QFT}

The field theory estimate proceeds from the oscillator discussion by considering many $a$'s: an $a_{k}$ for each spatial momentum $k$. In this simple case one simply adds the contribution for each $k$, so
\bea
| r_{\vec{k}} e^{i \theta_{\vec{k}}} \rangle=\prod_{k_{i}} \sum_{n_{i}=0}^\infty \frac{r^{n_{i}}_{k_{i}} e^{i n_{i} \theta}}{\sqrt{n_{i}}} \frac{(a^{\dagger}_{k_{i}})^{n_{i}} }{\sqrt{n_{i}} }| 0 \rangle=\prod_{k_{i}} e^{{r_{k_{i}}e^{{i\theta_{k_{i}}}}  a^{\dagger}_{k_{i}}}}|0\rangle\,.
\eea
We see that this is just a product of the coherent states of the previous case for each momentum $k_{i}$. If one writes $\rho_\beta=\prod_{k_i}\rho^{k_i}_\beta$ then
(\ref{thermalcoherent}) holds for each $\rho^{k_{i}}_\beta$ with $r\to r_{k_{i}}$ and $\theta \to \theta_{k_{i}}$ and therefore would also hold for its product, leading to the full thermal density matrix.

Its complexity, which is linear in the parameter that appears in the exponent, will be given by the sum of the individual complexities since
there does not seems to be a short cut even in this case:
\bea
\CC(| r_{\vec{k}} e^{i \theta_{\vec{k}}} \rangle, |0\rangle)=\sum_{k_{i}} r_{k_{i}}f(\theta_{k_{i}})\,.
\eea
Therefore
an upper bound on the mixed state complexity is given by
\bea
\Delta \CC_{E}^{b}=\sum_{k_{i}} r_{k_{i}}f(\theta_{k_{i}})p_{k_{i}}=\int \frac{d\theta f(\theta)}{2\pi } \sum_{k_{i}} \( \frac{e^{-\beta \omega_{k_{i}} }}{1-e^{-\beta \omega_{k_{i}} }}\)^{{1/2}}\,.
\eea

Consider a system of relativistic particles so $\omega_{k}\sim |k|$ (in this limit the theory becomes conformal), then going to the continuum one has
\bea
\Delta \CC_{E}^{b}\sim \frac{\textrm{V}\, \Omega_{D-2}}{(2\pi)^{D-1}}\int_{0}^{\Lambda} dk\,\frac{ k^{D-2}}{\sqrt{2}} \frac{e^{-\beta k/4}}{\sinh^{1/2}(\beta k/2)} =\frac{\sqrt{2}\,\textrm{V}\,\,\, T^{D-1}}{\pi^{\frac{D-1}{2}} \Gamma(\frac{D-1}{2}) } \int_0^\infty \frac{e^{-x/2} x^{D-2}}{\sinh^{1/2}x}dx  \nonumber \\
\eea
where $V$ is the spatial volume, $\Omega_{D-2}=2\pi^{(D-1)/2}/\Gamma((D-1)/2)$ is the volume of the $(D-2)$ dimensional sphere.  In the last integral we made the change of variables $x\to \beta k/2$, and noticed that the integral is finite for all values of $D>2$, so we removed the cut-off, taking $\Lambda \to \infty$. The numerical value of the integral can be evaluated case by case, for example for $D=3$ it takes the value $\pi \log(2)/\sqrt{2}$.

\section{Discussion}

This work analyzed various holographic proposals for subsystem complexity and compared results for eternal black holes to various qubit-based proposals for subsystem complexity. Simple tensor network models were used to develop intuition for the behavior of these measures in a strongly interacting system of qubits which might be expected to reasonably model the gross features of a holographic conformal field theory. While CA duality could be reasonably matched to the purification complexity or the basis complexity, we found that CV duality was somewhat in tension with the various proposals we considered. This tension arose in part because CV duality requires that subsystem complexities be superadditive with respect to the total system complexity. If the action is defined so that the UV divergent terms in CA are positive, then we found that the basis complexity is the best match for CA.

One interesting direction for future work is to search for other measures of complexity that might be better matched to CV duality; alternatively, one could try to modify CV duality, e.g., by including new contributions localized at the RT surface in the bulk. Another interesting direction is to try combining the notions of mixed state complexity studied here with more field-theoretic notions of pure state complexity. While we were able to draw some conclusions about the way different spacetime regions contributed to complexity, for example, the interior contribution in CA duality being associated with the spectrum preparation, there is still much more to learn about the way subregions influence the state complexity. It would also be very interesting to further explore holographic subsystem complexity in time-dependent situations, especially its covariant aspects.

\acknowledgments
We would like to thank Mohsen Alishahiha, Elena C\'aceres, Josiah Couch, Stefan Eccles, Hugo Marrochio, Reza Mozaffar, Rob Myers, and Phuc Nguyen for very useful discussions. We would also like to thank the anonymous referee for her or his many valuable suggestions on improving the presentation.

C.A. was supported in part by DOE grant de-sc0009987 and the National Science Foundation under CAREER awards PHY10-53842, PHY11-25915, and PHY16-20628.
M.H. was supported in this work in part by the NSF under Career Award No.\ PHY-1053842 and in part by the Simons Foundation through the ``It from Qubit'' Simons Collaboration and through a Simons Fellowshop in Theoretical Physics. M.H. would also like to thank the MIT Center for Theoretical Physics for hospitality while this research was undertaken.
B.G.S. is supported by the Simons Foundation through the ``It from Qubit'' Simons Collaboration and by the NSF under Grant No.\ NSF PHY-1125915. The authors would also like to thank the Kavli Insitute for Theoretical Physics at the University of California, Santa Barbara, where some of this work was undertaken.

\appendix

\section{Corner terms in subsystem complexity \label{AA}}

In the complexity equals action prescription it was argued that spacetime regions lying between null sheets give rise to additional boundary terms whose contribution to the action has the following form \cite{actionnull}
\beq
\CA_{\text{corner}} = \frac{1}{8 \pi G_N} \int_{\Sigma'} a\, dS
\eeq
where $\Sigma'$ is the corner locus (codimension two) and $a$ is the corner integrand given by
\bea
a = \pm \, \log \left|\frac{k \cdot \bk}{2} \right|
\eea
The sign depends on the causal relation between the region of interest (for the purpose of the action evaluation), the null sheets defining $\Sigma'$, and $\Sigma'$. The normals $k$, $\bk$ are 
the tangent vectors to the null sheet. Parametrizing the sheets by $\lambda = -\ell/z$ for a future boundary the region and $\lambda = \ell/z$ for a past boundary. 
Given the metric
\bea\label{metric}
ds^2=\frac{\ell^2}{z^2}\left(
-f(z)dt^2+\frac{dz^2}{f(z)} +\ell^2dx_{D-2}^2 \right),
\eea
the inner product between normals is therefore 
\bea
\half k\cdot \bk=-\frac{z^2}{f(z)}
\eea
and therefore for the corner terms we have
\bea \label{corner}
\frac{1}{8\pi G_{N}}\int_{\Sigma_{i}'} a dS=-\frac{\textrm{sgn}_i \, V_\perp}{8\pi G_{N}}\(\frac{\ell^2} {z_i}\)^{D-2}\log\(\frac{\ell^2\, f(z_i)}{z_i^2\, c \bc}\)\,,
\eea
where $z_i$ is the $z$ coordinate at the corner $i$.

In this appendix we study these corner terms as they appear in the subsystem complexity evaluation for the cases of charged and un-charged eternal black holes as the ones studied in sections (\ref{eternalCBH}) and (\ref{eternalBH}) which have precisely the form (\ref{metric}) with
\bea
f(z)=1-m \,z^{D-1} +q^2 \,z^{2(D-2)}
\eea
for the charged case, and
\bea
f(z)=1-(z/z_h)^{D-1}
\eea
for the uncharged one.

\begin{figure}
  \centering
  \includegraphics[width=\textwidth]{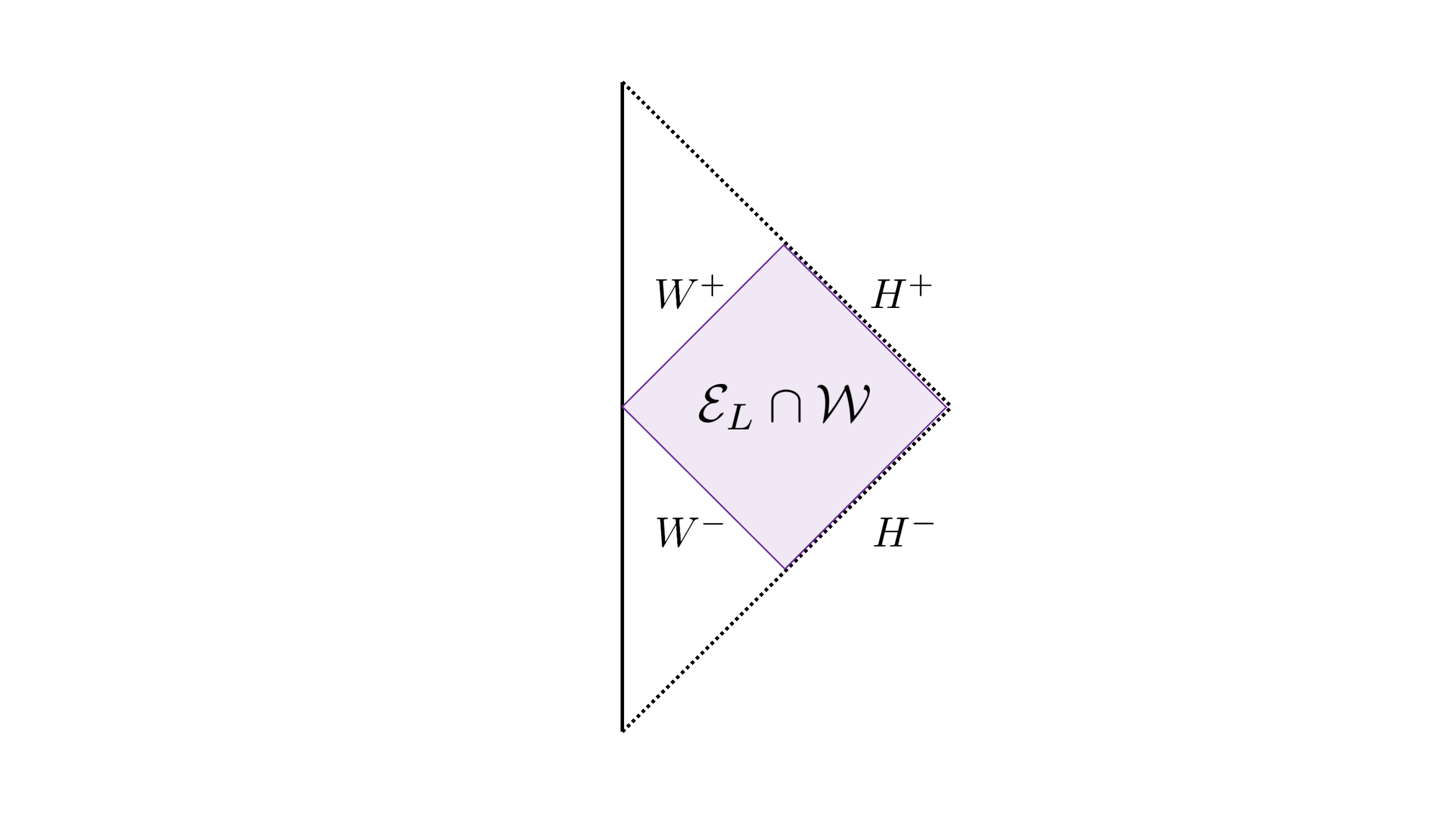}
  \caption{Null segments on the boundary of $\CW_L = \CE_L \cap \CW$. The four corners arise at the intersection of neighboring segments of the boundary. }\label{eternalbh_corners2}
\end{figure}

In that evaluation we have 4 corners: one on the boundary $W^{-} \cap W^{+}$ and three on the horizon $W^{+} \cap H^{+}$, $H^{+} \cap H^{-}$ and $W^{-} \cap H^{-}$ as illustrated in figure \ref{eternalbh_corners2}

A challenge posed by (\ref{corner}) is how to evaluate those expressions at the horizon, since they diverge at that exact location. Our strategy is to approximate the contribution from the surfaces lying on the horizon by approaching null surfaces. In that case the surface terms give zero contributions and the corners can be evaluated using (\ref{corner}). Once a regularized answer is obtained we expect to get a final finite result in the limit in which the surfaces approach the horizon. We will explain in detail such evaluation

First, let's consider a coordinate system in which the null surfaces are easily described. This can be achieved by the $u,v$  null coordinates defined as
\bea\label{nullcoor1}
u &=e^{f'(z_{h})(z^*(z)+t)/2}  \nonumber \\
v&=-e^{f'(z_{h})(z^*(z)-t)/2}
\eea
for $z<z_{h}$ left side of the exterior horizon region, and
\bea\label{nullcoor2}
u &=&e^{f'(z_{h})(z^*(z)+t)/2} \nonumber\\
v&=&e^{f'(z_{h})(z^*(z)-t)/2}
\eea
for $z>z_{h}$, which covers the future of the inside horizon region. Similar coordinate patches can be defined for the other two regions to cover the full geometry of the eternal black hole.

The coordinate $z^{*}(z)$ approaches a constant at the boundary and grows arbitrarily towards the horizon. If one fixes its boundary value to zero, this is $z^{*}(\delta)=0$ then the product
\bea\label{nullcoor}
uv=-e^{f'(z_{h})z^{*}(z)}, \quad \textrm{satisfies,}\quad uv=\begin{cases}
1 & \quad z=\delta , \\
0 & \quad z=z_{h} .
\end{cases}
\eea
since $f'(z_{h})$ is negative. We assume that $f'(z)$ is smooth and different than zero in a neighborhood of $z=z_{h}$; we will comment on the extremal charged case later on.

Since the left future  and past horizons $H^{+}$ and $H^{-}$ corresponds to the surfaces $v=0$ and $u=0$ respectively, we would like to consider instead the surfaces  $v=\ev$\footnote{ Here $\ev$ is negative.} and $u=\eu$ as the surfaces that approach them. Assuming that the boundary corner $W^{+}\cap W^{-}$ corresponds to the coordinate $(u_{0}, v_{0})$, then the three near-horizon corners will be at $(u_0,\epsilon_v)$, $(\eu,v_0)$ and $(\eu,\ev)$ which are associated to  three different radial coordinate points which we denote $z_{u_0, \ev},\, z_{\eu, v_0}$ and $z_{\eu, \ev}$ respectively. All these points have been regularized and depends on the regularization parameters $(\eu,\ev)$ as well as the physical information contained in $u_{0}$ and $v_0$.

Even though we don't know how to obtain the explicit function $z^{*}(z)$ for general $f(z)$ and then the relation between the $z$ coordinate and the parameters $u_0$ and $v_0$, one can find useful relations between them by using the equation
\bea
\log(-uv)=f'(z_h)z^*(z)
\eea
which is equivalent to (\ref{nullcoor}).

Notice that the way the corner contributions were derived in \cite{actionnull} was to guarantee the additivity property of the action formula. Therefore, it is interesting to check what does that imply to the regularization of the horizon surfaces as the one we proposed before. An obvious requirement is the following:

Consider a spacetime region that crosses the horizon at $v=0$ and divides the region in two on the horizon with corners at $u_1$ and $u_2$. The additivity property of the boundary tells us that the regularization should be such that the corner terms on opposite sides of the spacetime boundary regions cancel each other. That is, a given corner on both sides of the horizon comes with opposite signs and in principle, different regularized $v$ coordinates, which we called  $\epsilon_v$ and $\epsilon'_v$, where $\epsilon_v$ is the one associated to the surface outside the horizon and $\epsilon'_v$ the one behind the horizon. A pair of corner contributions for a given $u$ are proportional to
\bea
\log\(\frac{f(z_{u,\ev})}{z_h^2}\)-\log\(\frac{-f(z_{u,\ev'})}{z_h^2}\)
\eea
where both $z_{u,\ev}\approx z_h$ and $z_{u,\ev'}\approx z_h$. Then a reasonable requirement that makes this combination vanish would be
\bea
f(z_{u,\ev})=-f(z_{u,\ev'})
\eea
which implies a precised relation between $\ev$, $\ev'$.

However, this is a very strong and unnecesary constraint as the following manipulation shows. First, notice that the corners created in this way come in pairs so we can required instead the cancellation of the following combination
\bea\label{cornerss}
\log\(\frac{f(z_{u_1, \ev})}{z_h^2}\)-\log\(\frac{f(z_{u_2,\ev})}{z_h^2}\)-\log\(\frac{-f(z_{u_1, \ev'})}{z_h^2}\)+\log\(\frac{-f(z_{u_2,\ev'})}{z_h^2}\)\nonumber \\
\eea
and the difference of logarithms on a single side of the horizon can be simplified as follows:
\bea\label{ucut}
\log\(\frac{f(r_{u_1, \ev})}{z_h^2}\)- \log\(\frac{f(z_{u_2,\ev})}{z_h^2}\)&=&\int^{z_{u_1, \ev}}_{z_{u_2,\ev}} \frac{dz}{f(z)} f'(r)\approx f'(z_h)\int^{z_{u_1, \ev}}_{z_{u_2,\ev}} \frac{dz}{f(z)} \nonumber \\
&=&f'(z_h)(z^*(z_{u_1, \ev})-z^*(z_{u_2,\ev}))\nonumber \\
&=&\log(-u_1 \ev)-\log(-u_2\ev)
\eea
where we have used the fact that the derivative is smooth around the horizon. The result of (\ref{ucut}) is independent of the regularization, as long as the regulator is small and $z\approx z_h$. This makes the corner terms to cancel without the need of fine tuning the regulators outside and behing the horizons. Additionally, this constrains the form of the term
\bea\label{fuv}
\log\(\frac{f(z_{u,v})}{z_h^2}\)
\eea
as long as $z \approx z_h$. To derive the constraint, notice that when $v\approx -|\ev| \to 0$ the difference in (\ref{ucut}) has no divergence as can be seen from the RHS of that equation. In that limit the full dependence on $u$ of (\ref{fuv}) is logarithmic. Similarly, if one starts instead with a null surface $u=\eu$, with $\eu$ small one obtains the same conclusion, but now on the dependence on $v$. Combining these two limits one concludes that
\bea\label{flog}
\log\(\frac{f(z_{u,v})}{z_h^2}\)=\log(-uv)+F(z_h)
\eea
for $z\approx z_h$. So the function $F$ on the RHS of (\ref{flog}) which is cancelled in the difference (\ref{ucut}) has to be independent of both $u$ and $v$, so it would depend only on $z_h$. Furthermore, this means that in the $z\to z_h$ limit, the divergence structure of the LHS of (\ref{flog}) is completely given by $\log(uv)$ as $uv$ goes to zero.

One can obtain the value of the function $F(z_h)$ via
\bea\label{fzh}
F(z_h)&=&\lim_{z\to z_h}\(\log\(\frac{f(z)}{z^2}\)-f'(z_h)z^*(z) \)\,.
\eea
However, this computation requires the exact function $z^*(z)$ which we know only for the uncharged black hole of section (\ref{eternalBH}).

Nevertheless one can write a formal expression for the sum of the corner terms as
\bea
\sum_i\int_{\Sigma_{i}'} a_i dS_i&=&-\sum_i\textrm{sgn}_i \, V_\perp\(\frac{\ell^2}{z_i}\)^{D-2}\log\(\frac{f(z_i)}{z_i}\) \\
&=&-V_\perp\Bigg[{\(\frac{\ell^2}{z_h}\)}^{D-2}\Bigg\{\log\(\frac{f(z_{u_0,\ev})}{z^2_h}\)+\log\(\frac{f(z_{\eu,v_0})}{z^2_h}\)\nonumber \\
&& -\log\(\frac{f(z_{\eu,\ev})}{z^2_h}\)\Bigg\}-\(\frac{\ell^2}{\delta}\)^{D-2}\log\(\frac{f(\delta)}{\delta^2}\)\Bigg]
\eea
adding and substracting the corner term at $H^+\cap H^-$, one obtains two pair of corners which can be expressed in terms of differences of simple logarithms as in (\ref{ucut}), leading to
\bea
\sum_i\int_{\Sigma_{i}'} a_i dS_i
&=&-V_\perp\Bigg[{\(\frac{\ell^2}{z_h}\)}^{D-2}\Bigg\{\log\(-u_0 v_0\)+\log\(\frac{f(z_{\eu,\ev})}{z^2_h}\)-\log(-\eu \ev) \Bigg\}\nonumber \\
&&\qquad \qquad \qquad \qquad  -\(\frac{\ell^2}{\delta}\)^{D-2}\log\(\frac{f(\delta)}{\delta^2}\)\Bigg]\,.
\eea
In the $\eu,\ev\to 0$ limit we can recognize on the first line of the above equation the expression for $F(z_h)$. Similarly, doing a series expansion in small $\delta$ inside the $\log$ and keeping only the finite and divergent pieces one obtains
\bea\label{corners4}
\sum_i\int_{\Sigma_{i}'}a_i dS_i
&=&-V_\perp\Bigg[{\(\frac{\ell^2}{z_h}\)}^{D-2}F(z_h)+2\(\frac{\ell^2}{\delta}\)^{D-2}\log\delta\Bigg]
\eea
where we have used the condition $u_0v_0=-1$.

\subsection{Neutral black hole \label{nbh}}

For neutral BH one knows the explicit form of $z^*(z)$ and then the limit defining $F(z_h)$ can be computed exactly:
\bea
F(z_h)&=&\lim_{z\to z_h}\(\log\(\frac{f(z)}{z^2}\)-f'(z_h)z^*(z) \)\nonumber \\
&=&-2\log z_h+\lim_{z\to z_h}\left[\log\(1-\(\frac{z}{z_h}\)^{D-1}\)+B\(\(\frac{z}{z_h}\)^{D-1},\frac{1}{D-1},0\)\right] \nonumber \\
\eea
where we used $f'(z_h)z^*(z)=-B((z/z_h)^{D-1},1/(D-1),0)$.

The above limit can be evaluated by using the series expansion of $B(z/z_h,1/D-1,0)$, for $z<z_h$ which results in
\bea\label{logcc}
F(z_h)=-2\log z_h+\psi_0(1)-\psi_0\(\frac{1}{D-1}\)
\eea
where $\psi_0(z)$ is the digamma function defined as $\psi_0(z)=\Gamma'(z)/\Gamma(z)$, and $\psi_0(1)=-\gamma$ where $\gamma \approx 0.577216\dots$, is the Euler gamma constant.

The full  contribution of the corner terms for the subsystem complexity of neutral black holes can be written as
\bea\label{Finalcorners}
\frac{1}{8\pi G_N}\sum_i\int_{\Sigma_{i}'} a_i dS_i
&=&\frac{V_{\perp}}{4\pi G_N}\[- \( \frac{\ell^2}{\delta }\)^{D-2} \log \delta +  \( \frac{\ell^2}{z_h}\)^{D-2} \left(
g_0 + \log z_h \right)   \] \,,  \nonumber \\
\eea
where we have introduced the coefficient
\bea\label{g0}
g_0=\frac12 \left[\psi_0\(\frac{1}{D-1}\)-\psi_0(1) \right] \,
\eea
for ease of notation.

\subsection{Charged black hole}

Unlike for neutral black holes, for charged black holes we do not have an explicit expression for $z^*(z)$. Nevertheless, we know that the function $F(z_h)$ which in this case can depend on $z_h, q^2$ and $D$, is well defined and finite. Similar manipulations to the ones performed in \ref{nbh} are still of utility:
\bea
F(z_h)&=&\lim_{z\to z_h}\(\log\(\frac{f(z)}{z^2}\)-f'(z_h)z^*(z) \)\nonumber \\
&=&-2\log z_h+\lim_{z\to z_h}\left\{\log\(f(z)\)-f'(z_h)\int_0^z \frac{dz}{f(z)}\right\} \nonumber \\
\eea
which results in
\bea\label{Finalcorners2}
\frac{1}{8\pi G_N}\sum_i\int_{\Sigma_{i}'} a_i dS_i
&=&\frac{V_{\perp}}{4\pi G_N}\[ -\( \frac{\ell^2}{\delta }\)^{D-2} \log\delta +  \( \frac{\ell^2}{z_h}\)^{D-2} \left(
g(z_h) + \log z_h \right)   \] \,,  \nonumber \\
\eea
where
\bea\label{gzh}
g(z_h)=\lim_{z\to z_h}\left\{\log\(f(z)\)-f'(z_h)\int_0^z \frac{dz}{f(z)}\right\}\,.
\eea

\subsection{Extremal black hole \label{A3}}
The previous analysis only works for $f'(z_h)\neq0$, however it is well known that for extremal black holes this is not the case and so one would need to accommodate the previous procedure to account for those particular cases.

In these cases it is easy to modify the null coordinates (\ref{nullcoor1}), (\ref{nullcoor2}). The simple replacement $f'(z_h)\to -f''(z_h)z_h$ makes them a consistent coordinate chart. However, for this case there is no analogue of the relation between $z^*(z_{u,v})$ and $\log\(\frac{\ell^2 f(z_{u,v})}{z_h^2 c\bc}\)$ as the derived in (\ref{ucut}). The reason lies precisely in the fact that $f'(z_h)=0$. One question we would like to ask is whether there is an IR divergence in this case like the one obtained in the volume calculation coming from these corner terms.

We will do that in a perturbative fashion by expanding the functions $f(z_i)$ appearing in the evaluation of the corner contributions
\bea
\sum_i\int_{\Sigma_{i}'} a_i dS_i&=&-\sum_i\textrm{sgn}_i \, V_\perp\(\frac{l^2}{z_i}\)^{D-2}\log\(\frac{f(z_i)}{z_i}\) \\
&=&V_\perp {\(\frac{l^2}{z_h}\)}^{D-2} \Bigg[2\log z_h -\log\( f(z_{u_0,\ev}) \)-\log\( f(z_{\eu,v_0})\)\nonumber \\
&& \qquad \qquad \qquad \qquad +  \log\( f(z_{\eu,\ev})\)\Bigg] -2V_\perp {\(\frac{\ell^2}{\delta}\)}^{D-2}\log\delta\,.
\eea

In order to simplify the analysis we take the $\epsilon\to0$ limit in the following way. We chose $\epsilon_v=-\epsilon_u/u_0^2$ such that $u_0\epsilon_v=v_0\eu=-\eu/u_0$, which also implies $\eu\ev=-\eu^2/u_0^2$. In this case we have $z_{\eu,v_0}=z_{u_0,\ev}$. Since $z_{u,v}$ is actually a function of $(uv)$, to make this fact manifest we use instead $z_{u,v}=z(uv)$. In the $\eu\to 0$, $z\to z_h$ limit we have:
\bea
f(z)\approx \frac 12 f''(z_h)(z_h-z)^2
\eea
and therefore
\bea \label{IRdiv}
\sum_i\int_{\Sigma_{i}'} a_i dS_i&=& \lim_{\eu\to 0} 2V_\perp {\(\frac{\ell^2}{z_h}\)}^{D-2} \Bigg[\log z_h -\log\left\{ f\(z\(-\frac{\eu}{u_0}\)\)\right\} \nonumber \\
&&
\qquad \qquad \qquad +\frac 12 \log\left\{ f\(z\(-\frac{\eu^2}{u^2_0}\)\)\right\}\Bigg] -2V_\perp {\(\frac{l^2}{\delta}\)}^{D-2} \log\delta \nonumber \\
&\approx& \lim_{x\to 0} 2V_\perp {\(\frac{\ell^2}{z_h}\)}^{D-2} \Bigg[\log z_h -\frac 12 \log\left\{ \frac{f'' (z_h)z_h^2 }{2}\right\} \nonumber \\
&& \qquad \qquad \qquad -\log\left\{ \frac{(z_h-z(-x))^2}{z_h(z_h-z(-x^2))} \right\} \Bigg] -2V_\perp {\(\frac{l^2}{\delta}\)}^{D-2} \log\delta\nonumber \\
\eea
where $x=\eu /u_0$.
For these two $uv$ points, namely $uv=x$ and $uv=x^2$ we have the following relations
\bea
\log\(\frac{\eu}{u_0}\)&=&\log(x)=-z_h f''(z_h)\int_0^{z\(-x\)} \frac{dz}{f(z)} \nonumber \\
\log\(\frac{\eu^2}{u_0^2}\)&=&2\log(x)=-z_h f''(z_h)\int_0^{z\(-x^2\)} \frac{dz}{f(z)}
\eea
which came from the null coordinates (\ref{nullcoor1}), (\ref{nullcoor2}). We analyze only the leading divergence in the above equations in order to determine whether the full expression (\ref{IRdiv}) has a divergence as $x\to 0$. This is from the first relation we have
\bea
\log(x)&\approx &-z_h f''(z_h)\int_0^{z\(-x\)} \frac{dz}{f''(z_h)(z_h-z)^2/2} + {\rm subleading \,\, terms} \nonumber \\
&\approx &\frac{2z_h}{z_h-z(-x)}+ {\rm subleading \,\, terms}
\eea
while from the second one
\bea
\log(x)&\approx &-\frac{z_h f''(z_h)}2\int_0^{z\(-x^2\)} \frac{dz}{f''(z_h)(z_h-z)^2/2} + {\rm subleading \,\, terms}
\nonumber \\
&\approx &\frac{z_h}{z_h-z(-x^2)}+ {\rm subleading \,\, terms}  \,.
\eea
This means that at leading order, the divergences match with $(z_h-z(-x^2)) \sim (z_h-z(-x))/2$ and this in turn implies (after replacing that in (\ref{IRdiv})) that the corner terms in this case posesses an additional IR divergence. More precisely
\bea \label{IRdiv2}
\sum_i\int_{\Sigma_{i}'} a_i dS_i
&\approx& \lim_{x\to 0} 2V_\perp {\(\frac{\ell^2}{z_h}\)}^{D-2} \Bigg[\log z_h -\frac 12 \log\left\{ \frac{f'' (z_h)z_h^2 }{2}\right\} -\log\left\{ \frac{z_h-z(-x)}{4z_h} \right\}\nonumber \\
&& \qquad \qquad \qquad  \qquad \qquad \qquad + \quad {\rm subleading \,\, terms} \Bigg] -2 V_\perp {\(\frac{l^2}{\delta}\)}^{D-2} \log\delta \nonumber \\
\eea
which is logarithmically divergent as $x\to0$.

\section{Action calculations of pure-state complexity}
\subsection{Thermofield double \label{TFD}}

In this appendix we compute the pure-state complexity of the thermofield double state following the CA proposal.  In order to do so we will need to compute the action associated to the future and past interiors of the black hole considered in section (\ref{eternalBH}); these are the actions associated to the space-time regions denoted as $\mathcal{W}^{\pm}_{int}$.

The regions $\mathcal{W}^{\pm}_{int}$ are defined as the intersections of the future/past interiors of the black hole, and the WdW patch, respectively (the interior is the region between $z = z_h$ and the singularity, $z=\infty$).
The region  $\mathcal{W}^{-}_{int}$ is related to $\mathcal{W}^{+}_{int}$ by a time reflexion symmetry and therefore the action $\CA^{-}_{int}$ can be obtained from $\CA^{+}_{int}$ by doing a $t\to -t$ transformation. In what follows we will describe the calculation of $\CA^{+}_{int}$.

First notice that the tortoise coordinates $z^*(z)$ used in the light cone coordinates $u=\tau-z^*(z)$, $v=\tau+z^*(z)$ and defined in section (\ref{actioncal}) can be extended to cover the interior regions in a simple way. This is, for $z>z_h$ one can simply do
\bea
z^{*}(z)=z^*(z_h)+ \int_{z_h}^{z} \frac{dz}{f(z)}\,.
\eea
However, since in that region $f(z)$ is negative, the coordinate $z^*$ decreases as $z$ increases from $z=z_h$. To make the coordinate consistent, one then matches their values at $z=z_h$ and finds the relation between their asymptotics $z^*(\delta)$ and $z^*(\infty)\equiv z^*_{\infty}$. This is expressed in
\bea
z^*_\infty-z^*(\delta)=\int_{\delta}^{z_h} \frac{dz}{f(z)}+\int_{z_h}^{\infty} \frac{dz}{f(z)}\,.
\eea
The RHS integrals can be evaluated:
\bea
\int_{\delta}^{z_h} \frac{dz}{f(z)}= \frac{z_h}{D-1}\[ B\(1;\frac 1{D-1},0\)- B\(\delta_h^{D-1};\frac 1{D-1},0\)\]
\eea
from equation (\ref{tortD2}), and from (\ref{betaintegral}), one recognizes the result of the second integral after performing the change of variables $u=\xi^{1-D}$, this is
\bea
 \int_{z_h}^{z} \frac{dz}{f(z)}&=&\frac{z_h}{D-1}\int^{(z_h/z)^{D-1}}_{1} u^{\frac{D-2}{D-1}-1}(1-u)^{-1}du \nonumber \\
&=& \frac{z_h}{D-1}\[ B\(\(\frac{z_h}{z}\)^{D-1};\frac{D-2}{D-1},0\)- B\(1;\frac{D-2}{D-1},0\)\]\,. \label{zstar22}
\eea
 Finally, taken $\delta \to 0$ and $z\to \infty$ one gets
\bea\label{zstartdelta}
z^*_\infty-z^*(\delta)=\frac{z_h}{D-1}\[ B\(1;\frac 1{D-1},0\)- B\(1;1-\frac 1{D-1},0\)\]\,.
\eea
The right hand side quantity can be further simplified by writing the beta functions in a series expansion, and making use of the identity
\bea
\pi \cot \pi x=\frac 1x +\sum_{n=1}^{\infty}\frac{2x}{x^2-n^2}
\eea
where $x$ here is $1/(D-1)$. The equation (\ref{zstartdelta}) turns into
\bea
z^*_\infty-z^*(\delta)=\frac{\pi z_h}{D-1}\cot \(\frac{\pi}{D-1}\)\,.
\eea
This equations tells us that one cannot fix simultaneously both asymptotic values of $z^*$ for all $D\neq 3$. This is important in the construction of the Penrose diagram of Figure \ref{fig:eternalbh}.

\subsubsection{Calculation of $\mathcal{A}^{\pm}_{\rm int}$ \label{Aint+}}

Let's denote the initial time associated to the null rays that starts from the left boundary by $t_L$ and the parametric time that runs along the null line with $z$, $\tau$. In this case then we have the following:
\bea
\tau=t_L+\int_{\delta}^{z} \frac{dz}{f(z)}
\eea
In particular from this equation one can calculate $t_{c}$ (using the notation of \cite{actionnull}) the time at which the null ray hits the singularity exactly at $\tau=0$, this is
\bea
\tau=0=t_{c}+z^*_\infty-z^*(\delta)
\eea
and therefore:
\bea
t_{c}=z^*(\delta)-z^*_\infty=-\frac{\pi z_h}{D-1}\cot \(\frac{\pi}{D-1}\)
\eea
for $D\geq 3$.

That means that for initial times $t_{c}>t_L>t_{c}$ (which includes our case of interest $t_L=0$) the light ray will hit the singularity at a time $\tau$ of the interior geometry larger than zero and therefore the bulk integral of the full region ${ \cal W}^{+}_{{int}}$ including the singularity region is
\bea
\int_{\CW^{+}_{int}} dz d\tau\cdots=\int_{\infty}^{z_{h}}dz \int_{-t^R_{\infty}-(z^{*}(z)-z_{\infty})}^{t^L_{\infty}+z^{*}(z)-z_{\infty}}d\tau \cdots
\eea
where $t^{L/R}_{\infty}=t_{L/R}+z^{*}_{\infty}-z^{*}(\delta)=t_{L/R}-t_c$ and so replacing this back into the bulk integral
\bea
|{\cal W}^{+}_{int}|&=&-\hat{V}_{\perp} \ell^{2(D-1)} \Bigg[ \int_{\infty}^{z_{h}}\frac{dz}{z^{D}}(t_L+t_R-2t_c )  +\nonumber \\
&&\quad \qquad \quad + \frac{2 z_{h} }{(D-1)} \int_{\infty}^{z_{h}}\frac{dz}{z^{D}} B\( \(\frac{z_{h}}{z}\)^{D-1};\frac{D-2}{D-1},0\)\Bigg ]\nonumber \\
\eea
The first integral is trivial and the second one can be written in terms of the variable $x\equiv z_{h}/z$
\bea
|{\cal W}^+_{int}|&=&\frac{\hat{V}_{\perp}  \ell^{2(D-1)}}{(D-1)z_h^{D-2}}\[ \frac{t_L+t_R-2t_{c}}{z_{h}}+2\int_{0}^{1}x^{D-2}B\(x^{D-1};\frac{D-2}{D-1},0\)dx\] \nonumber \\
\eea
An extra change of variables for the remaining integral leads to
\bea\label{Integral2}
|{\cal W}^+_{int}|&=&\frac{\hat{V}_{\perp}  \ell^{2(D-1)}}{(D-1) z_h^{D-2}}\[ \frac{t_L+t_R-2t_{c}}{z_{h}}+ \frac{2}{D-1}\int_{0}^{1}B\(u;\frac{D-2}{D-1},0\)du \]\nonumber \\
\eea
where $u=x^{D-1}$. This is another member of the integrals computed in appendix (\ref{A}). Using equation (\ref{interiorint}), we get for the action
\bea
|{\cal W}^+_{int}|&=&\frac{2V_{\perp}  \ell^{2(D-1)}}{(D-1)(D-2) z_h^{D-2}} +\frac{V_{\perp}  \ell^{2(D-1)}}{(D-1) z_h^{D-1}}(t_L+t_R-2t_{c}) \,, \nonumber \\
\eea
and for the bulk contribution to the ``complexity''
\bea
{\cal A}^+_{\rm int,\, bulk}=-\frac{ V_{\perp} \ell^{2(D-2)}}{4\pi G_N (D-2) z_h^{D-2}}-\frac{ V_{\perp} \ell^{2(D-2)} }{8\pi G_N z_h^{D-1}}(t_L+t_R-2t_{c})\,.
\eea

In the interior region we also have a York-Gibbons-Hawing boundary term which gives a non-zero contribution on the space-like surface which covers the singularity, and zero on the light-like surfaces. The contributions has the form
\bea
\int_{\partial \mathcal{M}}\sqrt{|h|}K
\eea
where $K$ is the trace of the extrinsic curvature on $\partial \mathcal{M}$ and $h$ is its induced metric
\bea
K=\frac{n^{\mu}\partial_\mu(|h|)}{2|h|}\,,
\eea
with $n^{\mu}$ being the normal to the surface in question. At the $z=\infty$, we have $|h|=-l^{3D-4} f(z)z^{2-2D}$, and $\hat{n}=(z/l)\sqrt{-f(z)}\partial_z$ then
\bea\label{bdycontribution}
\frac{1}{16\pi G_N}\int_{z=\infty}\sqrt{|h|}K=\frac{ V_{\perp} (D-1)\ell^{2(D-2)}}{16\pi G_N z_h^{D-1}} (t_L+t_R-2t_c)\,.
\eea
We also have the counter term boundary contribution coming from the null surfaces that goes from the left and right black hole horizons to the singularity. Those terms can be evaluated as in section (\ref{actioncal}) leading to
\bea
 \frac{1}{8 \pi G_N} \textrm{sgn}(N)\int_{N} d\lambda d^{D-2}x \sqrt{\gamma} \Theta \log\(l_c |\Theta | \)&=&\frac{V_{\perp}\ell^{2(D-2)}}{4\pi G_N}\frac{\log \(l_c (D-2) z_h/\ell \)}{z_h^{D-2}}\nonumber \\
 &&+\frac{V_{\perp}\ell^{2(D-2)}}{4\pi G_N(D-2)}\frac{1}{z_h^{D-2}}
\eea

The corner calculation is a bit more involved although it follows directly from the procedure outlined in Appendix \ref{AA}. One has to be careful with the signs of the each corner as well as the sign change in $f(z)$. The result of that analysis leads to
\beq
{\cal A}^+_{\rm int,\, corners} = \frac{V_{\perp}}{8\pi G_N} \(\frac{\ell^2}{z_h}\)^{D-2}\left[ \log(u_L v_R)-2\log z_h -2g_0 \right].
\eeq
Here $g_0$ is the same that the one computed in Appendix \ref{AA}, (\ref{g0}) since the continuity imposed in $z^*(z)$ at $z_h$ guarantees it. The coordinates $u_L, v_R$ corresponds to the bounded light sheets equal to $u_L=\exp[-f'(z_h)(t_L-t_c)/2]$ and $v_R=\exp[-f'(z_h)(t_R-t_c)/2]$ and since they are computed in the region behind the horizon, $-f'(z_h)=(D-1)/z_h$ there, and then
\beq
{\cal A}^+_{\rm int,\, corners}= \frac{V_{\perp}}{8\pi G_N}\(\frac{\ell^2}{z_h}\)^{D-2} \left[ \frac{(D-1)}{z_h}\(\frac{t_L+t_R-2t_c}{2}\)-2\log z_h  -2g_0 \right]\,.
\eeq
Notice that for this range of boundary times $t_L,t_R$ the first term in the above expression is equal to the full term coming from the boundary contribution (\ref{bdycontribution}).

Adding all the contributions we get for the full interior action
\bea\label{actioninterior}
{\cal A}^+_{\rm int}&=&\frac{ V_{\perp} \ell^{2(D-2)}}{4\pi G_N z_h^{D-2} } \log\(\frac{l_c}\ell (D-2) \) -\frac{ V_{\perp} \ell^{2(D-2)}}{4\pi G_N z_h^{D-2} }g_{0}\nonumber \\
&& \qquad \qquad \qquad \qquad + \frac{ V_{\perp}(D-2) \ell^{2(D-2)} }{8\pi G_N z_h^{D-1}}(t_L+t_R-2t_{c})
\eea
in particular for $t_L=t_R=0$ we have
\bea\label{actioninterior-m2}
{\cal A}^+_{\rm int}&=&\frac{ V_{\perp} \ell^{2(D-2)}}{4\pi G_N z_h^{D-2} } \log\(\frac{l_c}\ell (D-2) \) -\frac{ V_{\perp} \ell^{2(D-2)}}{4\pi G_N z_h^{D-2} }g_{0} -\frac{ V_{\perp}(D-2) \ell^{2(D-2)} }{4\pi G_N z_h^{D-1}} t_c \,.\nonumber \\
\eea
Notice that for $\ell_c \geq \ell$ which was motivated in section (\ref{actioncal}) based on requiring the positivity of the divergent contribution in $\CC^{\rm A}_L(T)$, we find that each term in (\ref{actioninterior-m2}) is positive definite (since both $g_0$ and $t_c$ are negative) and therefore ${\cal A}^+_{\rm int}>0$.

For comparison purposes with the pure state complexity of the thermal field double state we would like to have also the past interior action in the same regime, this is $-t_c>t_L,t_R>t_c$. In that case the answer to that complexity contribution can be obtained from the future interior answer by simply doing $t_{L,R}\to -t_{L/R}$.

Recalling the additivity property of the action contribution one can write the pure state complexity in the interval $-t_c>t_L,t_R>t_c$ as
\bea
\CC^{\rm A}(\sigma) &=& \CC^{\rm A} (L)+\CC^{\rm A}(R)+\mathcal{A}^+_{\rm int}+\mathcal{A}^-_{\rm int}\nonumber \\
&=& \frac{ V_{\perp} \ell^{2(D-2)}}{2\pi G_N \delta^{D-2}} \log\(\frac{l_c}\ell(D-2) \)-\frac{ V_{\perp}(D-2)\ell^{2(D-2)}}{2\pi G_N z_h^{D-1}}\, t_c \nonumber \\
&=&\frac{ V_{\perp} \ell^{2(D-2)}}{2\pi G_N \delta^{D-2}} \log\(\frac{l_c}\ell(D-2) \)+2S \(\frac{D-2}{D-1}\)\cot \(\frac{\pi}{D-1}\)
\eea
where $\sigma=R\cup L$ is the full system. During that time interval the complexity is time independent as noted in \cite{Carmi:2017jqz}.

\subsection{Thermofield double at finite chemical potential}

In this appendix we would like to compute the pure-state complexity of the thermofield double state at finite chemical potential following the CA proposal. In order to do so we will need to compute the action associated to the future and past interior of the black hole considered in section (\ref{eternalCBH}), these are the space-time regions we denoted as $\mathcal{W}^{\pm}_{int}$.

As in the previous section, we only need to compute $\mathcal{W}^{+}_{int}$ since $\mathcal{W}^{-}_{int}$ is related to it by the time reflexion symmetry $t\to -t$. Additionally, the regions $\mathcal{W}^{\pm}_{int}$ do not intersect neither the future nor te past singularities and therefore our boundary surfaces will be all light-like in complete analogy with the computations of the actions associated to the regions $\mathcal{W}_{L/R}$ of section (\ref{eternalCBH}).

We start by properly delimiting the region $\mathcal{W}^{\pm}_{int}$. To do so we extend our tortoise coordinates to the region $z>z_h$ in order to cover the region behind the horizon, this is
\bea\label{doublestar}
z^{*}(z)\equiv z^*(z_h)+ \int_{z_h}^{z} \frac{dz}{f(z)}
\eea
where
\bea
 \int_{z_h}^{z} \frac{dz}{f(z)}=\int_{z_h}^{z}  \frac{dz}{1-m \,z^{D-1} +q^2 \,z^{2(D-2)}}\,.
\eea
As opposed to the previous case, here the light sheets coming from the left and right boundaries meet each other at a finite value of $z$ which we call $z_m$. This is so because the interior is the region between $z=z_h$ and $z=z_+$ (largest zero of $f(z)$).
Therefore the integral over the region covered by $\mathcal{W}_{int}^{+}$ is then
\bea
\int_{\CW^+_{int}} dz dt=-2\int_{z_{m}}^{z_{h}} dz \int_{0}^{z^{*}(z)-z^{*}(z_{m})} dt=-2\int_{z_{m}}^{z_{h}} dz\int_{z_{m}}^{z} \frac{d\xi}{f(\xi)}
\eea
An the bulk action calculation is therefore given by
\bea\label{actionsomething3}
{\cal A}^+_{\rm int, \, bulk}=\frac{(D-1)\hat{V}_{\perp} l^{2(D-2)}}{4\pi G_N }\int_{z_{m}}^{z_{h}} \frac{dz}{z^{D}}
\(1-\(\frac{D-3}{D-1}\) q^{2}z^{2(D-2)}\)\int_{z_{m}}^{z} \frac{d\xi}{f(\xi)} \,.
\nonumber \\
\eea
In this form, additional to the difficulty of carrying out the $\xi$ integral we encounter the problem of missing an explicit expression for $z_{m}$, which is implicitly given by
\bea\label{tLeft}
t_L+z^{*}(z_{m})-z^{*}(\delta)=0
\eea
where the right hand side time coordinate behind at the meeting point, which by symmetry it is zero. The initial times are assumed to be equal, that is $t_L=t_R$. Notice that the dependence on the initial time comes completely from $z_m$ via (\ref{tLeft}).

The difficulty in evaluating the bulk action is overcome by doing the same change in the order of integrals done in section (\ref{eternalCBH}) which leads to
\bea\label{actionsomething4}
{\cal A}^+_{\rm int, \, bulk}&=&\frac{(D-1)V_{\perp} \ell^{2(D-2)}}{4\pi G_N }\int_{z_{h}}^{z_{m}} \frac{d\xi}{f(\xi)} \int_{z_{h}}^{\xi} \frac{dz}{z^{D}}
\(1-\(\frac{D-3}{D-1}\) q^{2}z^{2(D-2)}\)\,.
\nonumber \\
&=&\frac{V_{\perp} l^{2(D-2)}}{4\pi G_N (D-2)}\(\frac{1}{z^{{D-2}}_{m}}-  \frac{1}{z^{{D-2}}_{h}}\)\,
\eea

In this case the boundary contribution comes only from the counter term, and can be evaluated as in section
 (\ref{actioncal}), with the difference that now the integral ends at the interesection point of the light-sheets, $z=z_m$, and then their contribution is
\bea
{\cal A}^+_{\text{int, boundary}}&=&\frac{V_{\perp}\ell^{2(D-2)}}{4\pi G_N(D-2)}\(\frac{1}{z_h^{D-2}}-\frac{1}{z_m^{D-2}} \)+\frac{V_{\perp}\ell^{2(D-2)}}{4\pi G_N}\frac{\log \(l_c (D-2)z_h/\ell \)}{z_h^{D-2}}\nonumber \\
&&-\frac{V_{\perp}\ell^{2(D-2)}}{4\pi G_N}\frac{\log \(l_c (D-2) z_m/\ell \)}{z_m^{D-2}}
\eea

The corner calculation is almost identical to the one described in \ref{Aint+} with an additional contribution coming from the intersection of the left and right light sheets at $z=z_m$, which results in
\bea
{\cal A}^+_{\rm int,\, corners} &=&
 \frac{V_{\perp}}{8\pi G_N} \(\frac{\ell^2}{z_h}\)^{D-2}\left[ \log(u_L v_R)-2\log z_h -2g(z_h) \right]
\nonumber \\
&& \qquad \qquad -\frac{V_{\perp}}{8\pi G_N} \(\frac{\ell^2}{z_m}\)^{D-2}\log\(-\frac{f(z_m)}{z_m^2}\)\,,
\eea
where $g(z)$ is given by (\ref{gzh}).  The coordinates $u_L, v_R$ corresponds to the bounded light sheets equal to $u_L=\exp[-f'(z_h)t_L/2]$ and $v_R=\exp[-f'(z_h)t_R/2]$, then
\bea
{\cal A}^+_{\rm int,\, corners}&=& \frac{V_{\perp}}{4\pi G_N}\(\frac{\ell^2}{z_h}\)^{D-2} \left[\frac{-f'(z_h)}{4}\(t_L+t_R\)-\log z_h -g(z_h) \right] \nonumber \\
&& -\frac{V_{\perp}}{4\pi G_N} \(\frac{\ell^2}{z_m}\)^{D-2}\left[ \log\(\frac{z_h \sqrt{-f(z_m)}}{z_m }\) -\log z_h\right]\,.
\eea
The full action associated to the region $\CW^+_{\rm int }$ at $t_L=t_R=0$ is thus
\bea
{\cal A}^+_{\rm int}&=&\frac{V_{\perp} \ell^{2(D-2)}}{4\pi G_N}\(\frac{1}{z^{{D-2}}_{h}}-  \frac{1}{z^{{D-2}}_{m}}\)\log \(\frac{l_c (D-2)}{\ell} \) \nonumber
\\
&& -\frac{V_{\perp}\ell^{2(D-2)} }{4\pi G_N}\left\{ \frac{g(z_h)}{z^{D-2}_h}  +\frac{1}{z^{D-2}_m}\log\(\frac{z_h \sqrt{-f(z_m)}}{z_m }\) \right\}\,.
\eea
Since $z_m>z_h$ the first line in the above equation is positive for $\ell>l_c$ which was previously motivated. However, the second line clearly positive as we do not have a direct handle on either $g(z_h)$ or $\sqrt{-f(z_m)}$. Nevertheless it is possible that ${\cal A}^+_{\rm int}$ goes from being positive to negative as one varies $\ell/l_c$ for given values of $m, q^2$.

Having all the pieces together we can easily obtain the complexity associated to the charged thermal field double state, using
\bea
\CC^{\rm A}(\sigma)&=& \CC^{\rm A}_L+\CC^{\rm A}_R+\mathcal{A}^+_{\rm int}+\mathcal{A}^-_{\rm int}
\eea
together with the fact that at $t_L=t_R=0$ we simply have $\mathcal{A}^+_{\rm int}=\mathcal{A}^-_{\rm int}$.
First, notice that the terms with explicit dependence on $z_h$ in $\mathcal{A}^{\pm}_{\rm int}$ cancel with the analogous terms in $\CC^{\rm A}_{L/R}$ and therefore
\bea\label{purecomplexitycharged}
C^{\rm A}(\sigma)&=& \frac{ V_{\perp} \ell^{2(D-2)}}{2\pi G_N} \log\(\frac{l_c}\ell(D-2) \)\[ \frac{1}{\delta^{D-2}}-\frac{1}{z_m^{D-2}}\]-\frac{V_{\perp}\ell^{2(D-2)} }{2\pi G_N}\frac{1}{z^{D-2}_m}\log\(\frac{z_h \sqrt{-f(z_m)}}{z_m }\) \,. \nonumber \\
\eea

\section{Integrals of incomplete beta functions\label{A}}

In the evaluation of the action associated to some space-time subregions we encountered integrals involving the incomplete beta function $B(z;a,b)$ given by
\bea\label{betaintegral}
B(z,a,b)\equiv \int_0^z dt (1-t)^{b-1}t^{a-1}dt\,.
\eea
This function reduces to the usual beta function when $z=1$, $B(z,a,b)=B(a,b)$ and has the following analytic expansion
\bea
B(z;a,b)=z^a\sum_{n=0}^{\infty} \frac{(1-b)_n}{n!(a+n)} z^n \,.
\eea
for arbitrary $z$. Here $(x)_n$ are the Pochhammer symbols.
When $b=0$ the above expansion simplifies to
\bea\label{seriesincbeta}
B(z;a,0)=\sum_{n=0}^{\infty} \frac{z^{a+n}}{a+n} \,,
\eea
since $(1)_n=n!$

The particular set of integrals we are interested in has the general form
\bea\label{intbeta}
\int x^{\beta}B(x;\alpha, 0)dx
\eea
which we aim to study here for positive and negative integer values of $\beta$.

Using the expansion (\ref{seriesincbeta}), one can rewrite the result of the indefinite integral  (\ref{intbeta}) as:
\bea\label{intbeta2}
\int x^{\beta}B(x;\alpha, 0)dx&=&\sum_{n=0}^{\infty} \frac{x^{\alpha + \beta +1+ n}}{(\alpha+n)(\alpha +\beta+1+n)}\nonumber \\
&=&\frac 1{\beta+1} \sum_{n=0}^{\infty} \frac{x^{\alpha+\beta+1+n}}{\alpha+n}-\frac 1{\beta+1}\sum_{n=0}^{\infty} \frac{x^{\alpha+\beta+1+n}}{\alpha+\beta+1+n}
\eea
up to an unimportant constant.

Now we will consider different cases of interest:

\paragraph{Case I:}
Consider $\beta$ to be a non-negative integer this is $\{ \beta \in \mathbb{Z}$, $\beta\geq 0 \}$ and assume the following relation holds $\{\alpha+\beta+1>0\}$ then
\bea\label{intbeta+}
\int x^{\beta}B(x;\alpha, 0)dx
&=&\frac 1{\beta+1} \sum_{n=0}^{\beta} \frac{x^{\alpha+\beta+1+n}}{\alpha+n}+\frac 1{\beta+1} \sum_{n=\beta+1}^{\infty} \frac{x^{\alpha+\beta+1+n}}{\alpha+n}\nonumber \\
&& \qquad \qquad -\frac 1{\beta+1}\sum_{n=0}^{\infty} \frac{x^{\alpha+\beta+1+n}}{\alpha+\beta+1+n}\nonumber \\
&=& \frac 1{\beta+1} \sum_{n=0}^{\beta} \frac{x^{\alpha+\beta+1+n}}{\alpha+n}+\frac 1{\beta+1}(x^{\beta+1} -1) B(x;\alpha+\beta+1,0)\, \nonumber \\
\eea
Since $\alpha+\beta+1>0$, all powers of $x$ are positive and one can evaluate the integral from  $x=0$ to $x=1$, leading to

\bea \label{intbeta+2}
\int_0^1 x^{\beta}B(x;\alpha, 0)dx
&=&\frac 1{\beta+1} \sum_{n=0}^{\beta} \frac{1}{\alpha+n}
\eea

\paragraph{Case II:}

Consider $\beta$ to be a negative integer whose magnitude is larger or equal to two, this is $\{ \beta \in \mathbb{Z}, -\beta \geq 2 \}$ and restrict $\alpha$ to be a non-integer, this is $\{ \alpha \notin \mathbb{Z} \}$. In that case then we have
\bea\label{intbeta-}
\int x^{\beta}B(x;\alpha, 0)dx
&=&-\frac 1{\beta+1} \sum_{n=0}^{|\beta|-2} \frac{x^{\alpha+\beta +1+n}}{\alpha+\beta+1+n}-\frac 1{\beta+1} \sum_{n=|\beta|-1}^{\infty} \frac{x^{\alpha+\beta+1+n}}{\alpha+\beta+1+n}\nonumber \\
&& \qquad \qquad +\frac 1{\beta+1}\sum_{n=0}^{\infty} \frac{x^{\alpha+\beta+1+n}}{\alpha+n}\nonumber \\
&=&-\frac 1{\beta+1} \sum_{n=1}^{|\beta|-1} \frac{x^{\alpha-n}}{\alpha-n}+\frac 1{\beta+1}(x^{\beta+1} -1) B(x;\alpha,0)\,.
\eea
Since the sign of $\alpha +\beta +1$ is undetermined we can not evaluate $x=0$ in general.
Instead we regularize the integral by evaluating it
from $x=\epsilon \approx 0$ to $x=1$ which leads to
\bea\label{intbeta-2}
\int_\epsilon^1 x^{\beta}B(x;\alpha, 0)dx
&=&-\frac 1{\beta+1} \sum_{n=1}^{|\beta|-1} \frac{1-\epsilon^{\alpha-n}}{\alpha-n}-\frac 1{\beta+1}(\epsilon^{\beta+1} -1) B(\epsilon;\alpha,0)\, \nonumber \\
\eea
Notice that in both cases we have used the fact that
\bea
\lim_{x\to 1}(x^{\beta}-1)B(x,\alpha,0)=0
\eea
for any $\beta$

In section \ref{actioncal} and \ref{TFD}, we have examples of these kind of integrals.
For example, in equation (\ref{Integral1}) we have
\bea\label{exteriorint}
\int_{\delta^{D-1}}^1 \frac{1}{u^2} B \left(u;\frac{1}{D-1},0 \right)du&=&-\( \frac{D-1}{D-2} \)\(1-\frac{1}{\delta^{D-2}}\)\nonumber \\
&&\qquad  +\(\frac{1}{\delta^{D-1                                   }}-1\) B \left(\delta^{D-1};\frac{1}{D-1},0 \right)
\eea
which belongs to the case II, and its value was obtained from (\ref{intbeta-2}) for
$\beta=-2$ and $\alpha=1/(D-1)$,
while in equation (\ref{Integral2}) we have
\bea\label{interiorint}
\int_0^1 B \left(u;\frac{D-2}{D-1},0 \right)du=\frac{D-1}{D-2}
\eea
which belong to the case I, and its value was obtained from (\ref{intbeta+2}) for  $\beta=0$ and $\alpha=(D-2)/(D-1)$.

\bibliographystyle{ucsd}
\bibliography{refs-subsystem}

\end{document}